\documentclass[
twocolumn,
superscriptaddress,
 amsmath,amssymb,
 aps,
prr,
floatfix,
longbibliography
]{revtex4-2}

\usepackage[utf8]{inputenc}
\usepackage[T1]{fontenc}

\usepackage{CJK}
\usepackage{multirow}
\usepackage{tabularx}
\usepackage{graphicx}
\DeclareGraphicsExtensions{.pdf,.png}
\usepackage{capt-of}
\usepackage[colorlinks=true, citecolor=blue, linkcolor=blue]{hyperref}

\begin{document}


\title{Self-limiting stacks of curvature-frustrated colloidal plates:  \\ Roles of intra-particle versus inter-particle deformations}

\author{Kyle T. Sullivan}
\affiliation{Department of Physics, University of Massachusetts, Amherst, Massachusetts 01003, USA}

\author{Ryan C. Hayward}
\affiliation{Department of Chemical and Biological Engineering, University of Colorado, Boulder, Colorado 80303, USA}

\author{Gregory M. Grason}
\affiliation{Department of Polymer Science and Engineering, University of Massachusetts, Amherst, Massachusetts 01003, USA}

\date{\today}

\begin{abstract} 
In geometrically frustrated assemblies local inter-subunit misfits propagate to  intra-assembly strain gradients, giving rise to anomalous self-limiting assembly thermodynamics.   Here, we use theory and coarse-grained simulation to study a recently developed class of ``curvamer'' particles, flexible shell-like particles that exhibit self-limiting assembly due to the build up of curvature deformation in cohesive stacks.  To address a generic, yet poorly understood aspect of frustrated assembly, we introduce a model of curvamer assembly that incorporates both {\it intra-particle} shape deformation as well as compliance of {\it inter-particle} cohesive gaps, an effect we can attribute to a {\it finite range of attraction} between particles.  We show that the ratio of intra-particle (bending elasticity) to inter-particle stiffness not only controls the regimes of self-limitation but also the nature of frustration propagation through curvamer stacks.  We find a transition from uniformly-bound, curvature-focusing stacks at small size to gap-opened, uniformly curved stacks at large size is controlled by a dimensionless measure of inter- versus intra-curvamer stiffness.  The finite range of inter-particle attraction determines range of cohesion in stacks are self-limiting, a prediction which is in strong agreement with numerical studies of our coarse-grained colloidal model.  These predictions provide critical guidance for experimental realizations of frustrated particle systems designed to exhibit self-limitation at especially large multi-particle scales.
\end{abstract}

\maketitle

\section{Introduction}

Geometric frustration occurs when a locally preferred ordering of the constituents of a system is unable to be achieved globally.  Originally associated with low-temperature magnetic spin ordering \cite{wannier-triangularising-1950,vannimenus-frustratedising-1977,collins-triangleantiferromagnets-1997}, geometric frustration has been studied in various condensed matter systems including colloidal ordering on curved surfaces \cite{meng-curvedcrystals-2014,irvine-pleatedcrystals-2010,guerra-freezingsphere-2018,li-crystalcaps-2019} and bent-core liquid crystals \cite{Reddy-BentcoreLC-2006,takezoe-bentcore-2006,fernandezrico-bananas-2020}.  In bulk systems, frustration is well-understood to result in the formation of  topological defects that localize the effects of shape mismatch \cite{kleman-curveddefects-1989}.  However, when featured in self-assembling systems, geometric frustration can lead to anomalous equilibrium morphologies and behavior, perhaps the most notable being finite assembly size \cite{grason-perspective-2016, meiri-gfa-2021}.  This paradigm of \textit{geometrically-frustrated assembly} (GFA) has been applied to understand different phenomena in soft matter including spherical protein shells \cite{mendoza-viralcapsids-2020}, twisted protein fibers \cite{aggeli-chiralrods-2001, turner-twistedproteins-2003, grason-chiralbundles-2007, hall-chiralbundles-2016, grason-twistedbundles-2020}, chiral ribbons \cite{ghafouri-ribbons-2005, zhang-ribbons-2019}, and assembled polyhedral nanoparticle mesostructures \cite{serafin-polyhedra-2021,schonhofer-polyhedra-2023, cheng_hyperbolic_2023}. 

In contrast with the simplest case of associating particles defined by either bulk, dispersed or defect-riddled condensed states, in GFA of soft matter systems there exists a possibilty of an intervening state of self-limiting assembly (SLA), in which equilibrium dimensions are finite but larger than subunit size~\cite{hackney-spins-2023}.  This self-limiting state relies on the super-extensive buildup of misfit strains and associated elastic costs over multi-particle dimensions~\cite{meiri-gfa-2021, grason-slareview-2021} that balances the energetic drive to bind additional attractive particles to select a finite assembly dimension. 

\begin{figure*}
    \centering
    \includegraphics[width=0.9\textwidth]{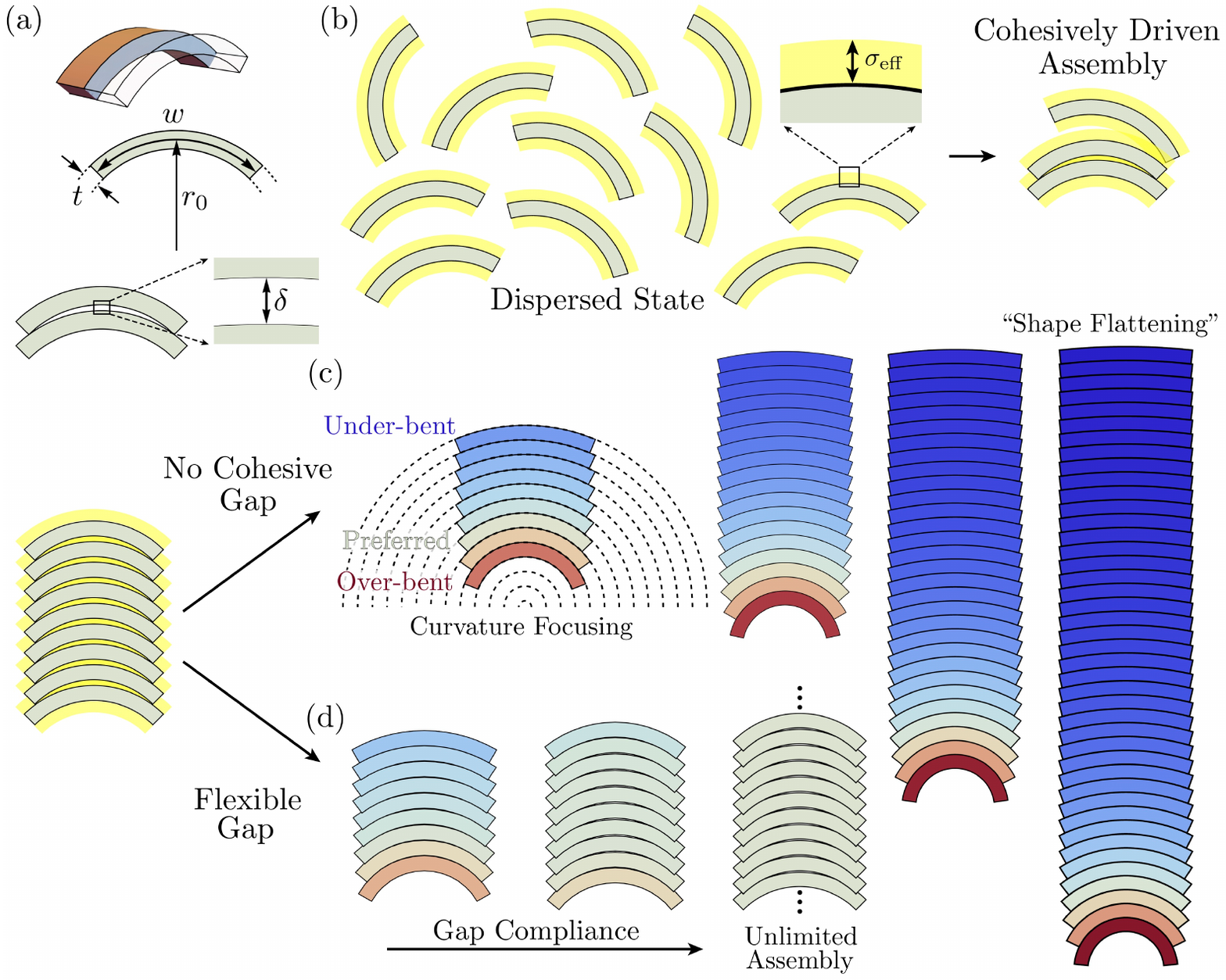}
    \caption{(a) The geometry of a cylindrical, shell-like ``curvamer'' in its preferred shape. A natural gap $\delta$ occurs between pairs of curvamers.  (b) Colloidal curvamers in a dispersed state attract and assemble into 1D stacks.  Yellow halos represent cohesive interactions between particles with an effective range of attraction $\sigma_{\rm eff}$.  (c) Curvamer assemblies with no cohesive gaps between particles, or equivalently interactions with $\sigma_{\rm eff} \ll \delta$, form concentric stacks of uniformly spaced, gap-closed particles whose curvatures are focused to a common focal point.  The buildup of curvature strains leads to finite-sized self-limiting states.  Large stacks can escape frustration by flattening the particles leading to unlimited growth.  (d) Assembly with flexible cohesive gaps, or equivalently finite attraction range, features both intra-particle (curvature change) and inter-particle (bond stretching) modes of elastic deformation.  Flexible gaps relieve stresses stemming from curvature change by allowing particles to be more uniformly shaped at the expense of opening gaps between particles.  Unlimited assembly may occur for extremely compliant gaps, or long attraction ranges ($\sigma_{\rm eff} \gg \delta$), as the particles become uniformly shaped and the center gaps uniformly stretched.}
    \label{fig:fig1}
\end{figure*}

The current understanding of self-limitation in GFA has relied primarily on continuum elastic frameworks of distinct models and has established a few basic principles \cite{grason-perspective-2016, grason-slareview-2021}.  First,  self-limiting size typically grows with the ratio of cohesion to elastic stiffness, while it {\it decreases} with increasing shape frustration.  Another theme is that for sufficiently soft frustrated systems, strong cohesive interactions cause the assembly to elastically ``defrustrate'' so that the each particle pays a constant misfit penalty to achieve an unfrustrated packing and unlimited assembly thermodynamics, dubbed a ``shape-flattening'' mode of frustration escape.  More recently, several discrete ``building block'' models have been introduced to study how microscopic features of subunits -- i.e. their misfit shapes, deformability and interactions -- control the range of physically accessible self-limiting assembly, and more generally engineer SLA behavior via fabrication of intentionally misfitting particle design~\cite{berengut-polybricks-2020}.  Notably, these discrete particle models so far fall into two distinct classes:  elastic polygons with infinitely short ranged interactions~\cite{lenz_polygon_17, leroy_collective_2023, botond-tubules-2022} or rigid, shape-frustrated particles with finite-ranged attractive interactions~\cite{Spivack-puzzlemers-2022, hall-wedges-2023}.  In the former case, elastic costs of frustration are borne entirely by subunit deformation, while in the latter misfit strain leads to stretching of inter-particle attractions.  In physical particle assemblies, the costs of frustration will be distributed to a combination of inter-particle {\it and} intra-particle deformation, according to the minimal free energy state for a given aggregate state.  Though at present, it remains poorly understood what controls when which deformation mode dominates, and more importantly, what are the distinct consequences for assembly thermodynamics should frustration strain be accommodated in one or the other.


A recently developed model of discrete GFA subunit, introduced by Tanjeem and coworkers \cite{conformalcurvamers} suggests that the interplay between intra- and inter-particle deformations has important implications for the range of self-limiting assembly behavior.  This model considers a stacking assembly of a cylindrical, shell-like, colloidal particle model (dubbed ``curvamer''), shown schematically in Fig. \ref{fig:fig1}.  Here, a simplified continuum theory for curvamer stacks was developed based on the  assumption of perfectly-contacting binding geometries favored by strong cohesive interactions.  In this perfect-contact approximation, it was assumed that the intra-assembly stresses were solely borne out of the particles changing curvature, due to a so-called ``curvature focussing'' effect required by uniform spacing of curved layers~\cite{sethna_smectic_1982, didonna_blue_2003}.  This perfect-contact model predicted that self-limiting stack sizes are possible for {\it arbitrarily} large cohesive forces between the particles. However, comparison to simulations of a discrete, coarse-grained model of curvamer stacks in the same study found that self-limiting stacks are only favorable over unlimited stacks  for sufficiently weak cohesive binding.  Furthermore, the range of self-limiting assembly was shown to {\it decrease} with the range of cohesive forces between curvamers.  This observation, along with the apparent deviation of discrete curvamer simulations away from perfect-contact, particularly, with longer range interactions, suggested that cohesive strain of the inter-curvamer binding itself is crucial for understanding thermodynamically optimal states of this class of frustrated assembly.  

We can understand this schematically by comparing the range of cohesive interactions, $\sigma_{\rm eff}$ (shown as a yellow halo surrounding curvamers in Fig. \ref{fig:fig1}), to the size of the natural gap, $\delta$, between curvamers which maintain their preferred shape.  When $\sigma_{\rm eff} \ll \delta$, particles must adjust their shapes to a curvature focusing geometry in order to close the gap and gain cohesive energy.  Alternatively, when $\sigma_{\rm eff} \gg \delta$, curvamer surfaces favorably attract over their entire width without changing shape to close the gap.  Heuristically, this suggests a transition from strongly-bound, curvature focused stacking to gap-opened, uniformly shaped stack geometries with increasing range and compliance of cohesive forces.  This picture raises a number of key questions about the impact of compliant interactions in frustration limited assembly of curvamers.  What physical parameters govern the distribution of frustration induced stress to either particle deformation or inter-particle bond stretching?   How do these distinct deformation modes control the accumulation of self-limiting elastic energies in curvamer stacks?  When interactions are compliant and finite ranged, what range of self-limiting stack assembly is possible and how is this controlled by shape, elastic and interaction parameters for a colloidal curvamer particle? 


To address these questions, here we extend and analyze the continuum theory of curvamer assembly introduced in Ref. \cite{conformalcurvamers} to incorporate both intra-particle (bending) and inter-particle (bond) deformations.  In parallel, we compare numerical studies of optimal stacking in a coarse-grained model of discrete colloidal curvamers.  We analyze the ground state stack energetics as a function of stack size as well as a  new parameter measuring the ratio of inter- vs. intra-curvamer stiffness.  This analysis predicts a generic transition from curvature-focusing to gap-opened stacking configurations that is controlled by a combination of size and relative compliance of interactions and particle shape.  Most significantly, deformable interactions modified the structure and energetics of asymptotically large stacks, limiting the accumulation of elastic energy due to shape misfit.
This, in turn, can be directly connected to a second-order like transition from self-limiting to unlimited curvamer stacking that occurs at a critical cohesive strength, which increases with the ratio of cohesive to shape stiffness.  
Based on this model, we construct the phase diagram of self-limiting vs. unlimited stack assembly and show that it is controlled by dimensionless measures of cohesive strength and range of interaction.  Crucially, we predict that self-limiting assembly is only possible below a maximal range of attractive forces between colloidal curvamers.  These predictions provide necessary guidance for the experimental design and study of attractive and curved colloids, from banana shaped-particles to lithographically fabricated polymeric shells, specifically for efforts towards realizing ``programmable'' large-scale assemblies of frustrated particle systems.

The remainder of this article is organized as follows.  In Sec. \ref{section:models} we introduce both our continuum analytic theory and coarse-grained simulation model for assembly of curvamers with flexible interactions.  Next, we analyze energetic ground state structures in mechanical equilibrium and describe the transition from curvature-focusing to gap-opened stacking in Sec. \ref{section:structures}.  Then in Sec. \ref{section:sla} we assess the range of accessible self-limiting stacking behavior in terms of a finite ratio of inter-particle stiffness to intra-particle bond stiffness, and alternatively in terms of finite ranges of attraction ultimately providing a phase diagram of self-limiting versus unlimited assembly that depends on attraction range and cohesive strength.  Finally, we summarize our results and discuss the implications for different possible experimental designs of self-limiting stacks of flexible, curved colloidal particles in Sec. \ref{section:discussion}. In particular we discuss systems of banana colloids and polymeric shell particles that interact through depletion forces, as well as DNA origami nanostructures that bind using single stranded DNA hybridization.    


\section{Models of Curvamer Stacking Assembly}
\label{section:models}

Starting from the model of conformal curvamer assembly presented in Ref. \cite{conformalcurvamers}, we model each curvamer as a two-dimensional, curved shell with thickness $t$, mid-line width $w$, and preferred radius of curvature $r_0 = \kappa_0^{-1}$, as shown in Fig. \ref{fig:fig1}.  While curvamer particles may be realized by cylindrical shells stacking in three-dimensions, the models introduced here focus on a simpler two-dimensional picture based on the assumption that optimal binding geometries favor maximal overlap between attractive surfaces, allowing us to focus on the cross-sectional shapes, which appear as one-dimensional curved bars of finite thickness.  Since deformations are assumed to be constant in these one-dimensional cross-sections, the elastic energy derives purely from {\it bending energy} away from the preferred curvature  $\kappa_0$.   We assume that the concave ``bottoms'' of particles favorably bind to convex ``top'' surfaces of adjacent curvamers and that this binding is mediated by a finite-ranged, surface-to-surface attractive force.  Following Ref. \cite{conformalcurvamers} we assume that the predominant state of stacking assembly maintains curvamer alignment in the stack, which permits our theoretical analysis of the elastic energy of stacks of variable size.  Notably, as reported in Ref. \cite{conformalcurvamers}, under certain cases, aligned curvamer stacks may be unstable with lateral-sliding instabilities that give rise to more complex (i.e. non-mirror symmetric) stacking geometries.  However, as we exploit below, lateral motion of bound curvamers can be suppressed through the introduction of patchy interactions in which attractive zones are confined to the central regions of curvamer faces.  Hence, in the analysis of curvamer assembly thermodynamics here, we neglect the possibility of these more complex, misaligned stacking motifs.

\subsection{Continuum mechanical model of curvamer stacking}
\label{section:continuummodel}

Here we construct an analytical model of $N$ consecutive bound curvamers in a stack, accounting for both the mechanics of shape deformation {\it and} inter-particle bond stretching.  For simplicity, we assume that curvature along a given particle is constant to good approximation, so that every section is a circular arc.  When two curvamers at their preferred curvature stack on top of one another without overlap, there is a natural gap $\delta\approx \frac{1}{8} t \kappa_0 ^2 w^2$ between their centers (Fig. \ref{fig:fig1}a).  In general, attractive interactions favor closing that gap, which requires deformation of particle shape.  Assuming these curvamers are sufficiently thin to neglect extension of their length, we model this via an elastic shell energy

\begin{equation}
E_{\text{bend}}(n) = \frac{1}{2} B w (\kappa_n - \kappa_0)^2\text{,}
\end{equation}    
where $\kappa_n$ is the (mid-line) curvature of the $n^{\rm th}$ particle in the stack and $B$ is the bending modulus, so that curvamers experience a linear force response for deviations away from the preferred curvature \footnote{Note that this bending modulus $B$ is the modulus of the {\it cross-section} of a curvature and has units of energy times length, as in beam bending.  Relating to the flexural modulus of a 2D shell, therefore requires mutliplication by the uniform, lateral size of the shell.}. 

\begin{figure}
    \centering
    \includegraphics[width=0.45\textwidth]{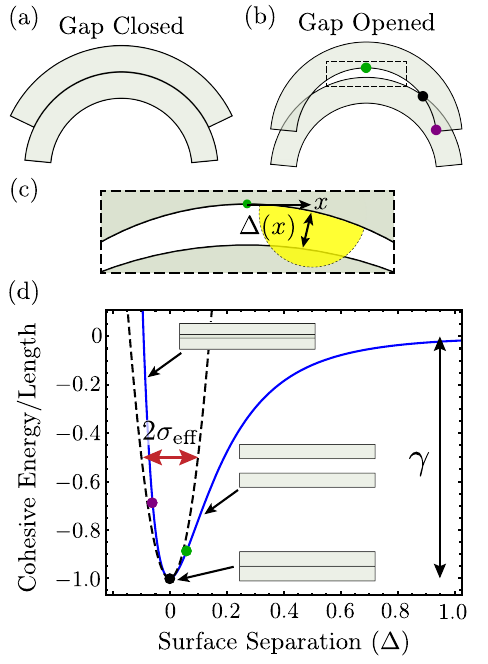}
    \caption{(a) Curvature focusing particles stack with perfect contact.  (b) Non-curvature focused particles stack with a gap that varies along their surfaces.  The gap is stretched open in the middle of the particle (green dot), compressed towards the flanks (purple dot) and is zero at a point in between (black dot).  (c) Close up view of the gap between non-curvature focused particles.  The gap distance $\Delta(x)$ is measured at a point $x$ away from the center of the top particle.  (d) The cohesive interaction energy per unit length between curvamer surfaces as a funtion of surface-surface separation distance.  The interaction is approximated as a harmonic well around its minimum.  An effective interaction range is defined as $\sigma_{\rm eff} = \sqrt{\gamma/\gamma''}$, or half the width of the well at half its minimum.}
    \label{fig:harmonicapprox}
\end{figure}

For the extreme case of infinitely-short ranged attraction, inter-particle binding requires an assembly of curvamers to stack conformally (without gaps), and the radius of curvature of the $n^{\text{th}}$ particle must change concentrically away from the radius of curvature of the bottom curvamer in the stack as $r_n = r_{n-1} + t$.  If this conformal stacking motif propagates through the entire assembly, we have a shape profile $r_n = r_{-} + nt$, or 

\begin{equation}
\label{equation:curvefocus}
    \kappa_n = \frac{\kappa_{-}}{1 + n\kappa_{-} t} \approx \kappa_{-} - n\kappa_{-}^2 t , \ ({\rm curvature \ focusing})
\end{equation}
where $\kappa_{-}$ is the curvature of a ``virtual'' curvamer at the bottom the stack ($n=0$).  We refer to this as \textit{curvature focusing}, due to the divergence of the curvature at the center of the concentric packing  (a.k.a. the focal point) and can be seen schematically in Fig. \ref{fig:fig1}c.  The ``gradient'' of shape in a conformal stack, and its associated elastic energy, grows with stack size and provides the fundamental mechanism to compete with cohesive drives to bind additional particles in a stack.

To introduce the additional possibility of non-conformal assembly (gap-opened stacking), we define a cohesive energy per unit length between pairs of curvamers 

\begin{equation}
\frac{dE_\text{coh}}{dx} = -\gamma + \frac{1}{2}\gamma'' \Delta^2(x)\text{,}
\end{equation}
where $\Delta(x)$ is the surface-surface separation (the distance of closest approach) at a point $x$ away from the center of the particle.  Interactions of this form, which treat the composite interaction as the superposition of locally planar geometries, are the generic consequence of finite-range colloidal forces between surface elements, which themselves derive from the combination of short-range repulsive forces that prevent overlap and long-range attractions \cite{israelachvili_intermolecular_1992}.  Hence, local interactions are described by a local equilibrium spacing, which we describe by surfaces in contact, and a finite range of interaction that describes distortion away from the local minimum (see Fig. \ref{fig:harmonicapprox}).  In our analytic model, we parameterize the effect of deformations away from the locally preferred spacing by an energetic penalty $\gamma''$ for opening gaps between curvamers.  As shown in Fig. \ref{fig:harmonicapprox}, this corresponds to a harmonic approximation of interactions around their local minimum, which is strictly accurate, provided distortions are small enough compared to the effective range of interactions $\sigma_{\rm eff} = \sqrt{\gamma/\gamma''}$.  Intuitively, as $\sigma_{\rm eff}$ decreases, cohesive interactions become effectively {\it more stiff}.

We note that when the two curvamers feature curvature focusing, the surface-surface gap becomes a constant ($\Delta(x) = \Delta$) and the total cohesive energy can be minimized to $E_{\text{coh}}=-\gamma w$ when $\Delta=0$, meaning the curvamers are stacked conformally as seen in Fig. \ref{fig:harmonicapprox}a.  If the pair is not curvature focusing, however, the surface-surface separation $\Delta(x)$ is not constant and will depend on the deviation of $\kappa_n$ and $\kappa_{n+1}$ from a curvature focusing configuration.  To calculate $\Delta(x)$, we assume $\kappa w \ll 1$ allowing us to approximate the circular curvamers as parabolas and find
\begin{equation}
    \Delta(x) \simeq \Delta z - t - \frac{1}{2} \left(\kappa_{n+1}^{-} - \kappa_n^{+}\right)x^2,
    \label{eq: gapstrain}
\end{equation}
where $\Delta z$ is the center-center separation distance between curvamers, and $\kappa_{n}^{\pm}$ are the curvatures of the top ($+$) and bottom ($-$) surfaces of the $n^{\text{th}}$ curvamer with midline curvature $\kappa_n$.  Averaging the square gap-strain over the curvamer width we find
\begin{multline}
    \langle \Delta^2(x) \rangle 
    = (\Delta z - t)^2 -\frac{w^2}{12}\left(\kappa_{n+1}^{-} - \kappa_n^{+}\right)(\Delta z - t) \\ +\frac{w^4}{320}\left(\kappa_{n+1}^{-} - \kappa_n^{+}\right)^2\text{,} 
\end{multline}
which is minimal for a center-to-center spacing $\Delta z^* = t +\frac{1}{24}\left( \kappa_{n+1}^{-} - \kappa_n^{+} \right)w^2$.  Notably, this optimal vertical spacing is only equal to particle thickness when particles are curvature focusing, i.e. $\kappa_{n+1}^{-} = \kappa_n^{+}$, and the gap strain consequently vanishes.  From this optimal spacing, we find the (width-average) cohesive energy to be

\begin{align}
    & E_{\text{coh}}^{n,\,n+1} (\Delta z^*) = -\gamma w + \frac{1}{2} \frac{\gamma'' w^5}{720}\left( \kappa_{n+1}^{-}-\kappa_n^{+} \right)^2\\
    & \simeq -\gamma w + \frac{1}{2} \frac{\gamma'' w^5t^2}{720}\left( \frac{\kappa_{n+1}-\kappa_{n}}{t}+\frac{\kappa_{n+1}^2+\kappa_n^2}{2}  \right)^2\text{,}
\end{align}
where in the second line we assume $\kappa t \ll 1 $ to approximate the top and bottom curvatures as $\kappa_n^{\pm} = \frac{\kappa_n}{1 \pm t\kappa_n /2} \simeq \kappa_n \mp \frac{1}{2}t\kappa_n^2$.

Defining the {\it reduced curvature} as $\tilde{\kappa} = \kappa/\kappa_0$, we then find the total energy of a stack of $N$ curvamers and normalize by $Bw\kappa_0^2$, an energy scaling characterizing curvature flattening,
\begin{align}
\label{equation:ediscrete}
    & \frac{E_{\rm stack}}{Bw\kappa_0^2} = -\frac{\gamma}{B\kappa_0^2}(N-1)+\frac{1}{2}\sum_{n=1}^{N}\left( \tilde{\kappa}_n - 1\right)^2 \nonumber \\ & +  \frac{1}{2}\frac{\gamma'' w^4 t^2 \kappa_0^2}{720 B}\sum_{n=1}^{N-1}\left( \frac{\tilde{\kappa}_{n+1} - \tilde{\kappa}_n }{t\kappa_0} + \frac{\tilde{\kappa}_{n+1}^2 + \tilde{\kappa}_n^2}{2} \right)^2 \text{.}
\end{align}
We next take the continuum limit of this stacking energy in the limit $N\gg1$ and $t\kappa_0\ll 1$, where the sums in Eq. (\ref{equation:ediscrete}) are well-approximated as integrals and likewise $\kappa_{n+1} - \kappa_n \approx \partial \kappa / \partial n$.  It is convenient to reparameterize the position in the stack by the scaled height coordinate
\begin{equation}
    h \equiv n (\kappa_0 t)\text{,} 
\end{equation}
and define the dimensionless ratio of cohesive energy relative to intra-particle stiffness
\begin{equation}
    S = \frac{\gamma t}{B\kappa_0}\text{.}
\end{equation}
Notably, this latter quantity can be understood as the ratio of the ``surface energy'' of the missing cohesion at the top and bottom of the stack, $\gamma w$, relative to the cost of flattening a stack of curvamers of thickness equal to $Nt = r_0=\kappa_0^{-1}$, given by $Bw\kappa_0/t$.  As shown in Ref. \cite{conformalcurvamers}, $S$ controls the equilibrium self-limiting stack size for the conformal limit of curvamer assembly.  

As self-limiting thermodynamics is controlled by the size dependence of (interaction free) energy per particle, we define the dimensionless total energy density in the stack of scaled size $H = N (\kappa_0 t)$,
\begin{align}\label{equation:econtinuum}
    \epsilon (H) \equiv \frac{E/N}{Bw\kappa_0^2} = - \epsilon_0 + \frac{S}{H} + \epsilon_{\rm ex} [ \tilde{\kappa}(h)]\text{,}
\end{align}
where $\epsilon_0 = S/(\kappa_0 t)$ is a measure the bulk cohesive energy of assembly while $S/H$ is the per particle cost of the unbound surfaces at the top and bottom of the stack.  The {\it excess energy} density, $\epsilon_{\rm ex} [ \tilde{\kappa}(h)]$, represents that accumulation of additional costs of assembly associated with the frustration~\cite{grason-slareview-2021}, which here depends on the shape profile of curvature $ \tilde{\kappa}(h)$ in the stack,
\begin{equation}
\label{equation:eexcess}
\epsilon_{\rm ex} [ \tilde{\kappa}(h)] = \frac{1}{H} \int_0^H \left[ \frac{1}{2} (\tilde{\kappa} -1)^2 + \frac{G}{2} (\tilde{\kappa}' +\tilde{\kappa}^2)^2   \right] {\rm d}h ,
\end{equation}
where $\tilde{\kappa}' = \partial \tilde{\kappa} / \partial h$ and we introduce the dimensionless quantity
\begin{equation}
G = \frac{\gamma'' w^4 t^2 \kappa_0^2 }{ 720 B}\text{,}
\end{equation}
that parameterizes the cohesive (inter-particle) stiffness to bending (intra-particle) stiffness ratio.  This can be understood (up to a prefactor) as the ratio of the energetic cost to stretch inter-particle bonds between parallel surfaces a distance $\delta$, given by $\gamma''w\delta^2 \sim \gamma''t^2w^5\kappa_0^4$, to the flattening energy of a stack of thickness $Nt = r_0$.  Hence, the first term in the functional favors uniform, preferred  shape (i.e. $\tilde{\kappa} =1$), while the latter term parameterizes the cost of gap strain, vanishing only for curvature focusing profiles (i.e. $\tilde{\kappa}' = -\tilde{\kappa}^2$ in the continuum limit).  Predicting the accumulation of excess energy with stack size $H$, and relating that to self-limiting thermodynamics requires optimizing the functional $\epsilon_{\rm ex} [ \tilde{\kappa}(h)] $ with respect to curvature profile in the stack.  For a given size, optimal energy stacks satisfy {\it mechanical equilibrium} described by solutions of the Euler-Lagrange equation
\begin{equation}
\frac{d}{dh} \left[\frac{G}{2} (\tilde{\kappa}')^2 +\frac{1}{2} (\tilde{\kappa}-1)^2 +\frac{G}{2} \tilde{\kappa}^4\right] = 0\text{,}
\label{eq: ELeq}
\end{equation}
subject to the free boundary conditions 
\begin{equation}
\tilde{\kappa}'(0) =-\tilde{\kappa}^2(0); \ \tilde{\kappa}'(H) =-\tilde{\kappa}^2(H)\text{,}
\label{eq: BCs}
\end{equation}
which means that stacks satisfy curvature focusing at the open boundaries on their top and bottom.  Detailed solutions for these equations are provided in Appendix \ref{appendix:curvprofile}, and take the form of elliptic integrals well known for the Euler elastica problem satisfying the same class of non-linear ODE~\cite{elastica}.  Here, we briefly remark that combination of Eqs. (\ref{eq: ELeq}) and  (\ref{eq: BCs}) gives a surprisingly simple relation between curvature of the two open boundaries of mechanically equilibrated stacks
\begin{equation}
\big| \tilde{\kappa}_-- 1\big|^2 = \big| \tilde{\kappa}_+- 1\big|^2 ,
\label{eq: overunder}
\end{equation}
where $\kappa_- \equiv \kappa (0)$ and $\kappa_+ \equiv \kappa (N)$ denote the curvature at the bottom and top of the stack respectively.  Hence, this variational model makes generic predictions for ends of stacks in mechanical equilibrium for any size and notwithstanding their top vs. bottom asymmetry:  (i) stacks approach curvature focusing at their free boundaries and (ii) the amount of ``overbending'' on one end of the stack is equal to the degree of ``underbending'' on the opposite end.  We denote the (energy minimizing) mechanical equilibrium solutions of Eqs. (\ref{eq: ELeq}) and  (\ref{eq: BCs}) as $\tilde{\kappa}_* (h)$ and will describe their structure in Sec. \ref{section:structures} below.  We discuss the thermodynamic dependence of (profile optimized) excess energy $\epsilon_{\rm ex} (H, G) = \epsilon_{\rm ex} \left[ \tilde{\kappa}_* (h) \right]$ on stack size in Sec. \ref{section:structures} and its implications for self-limiting stack formation in Sec. \ref{section:sla}.


\subsection{Discrete, coarse-grained colloidal model}

\begin{figure}
    \centering
    \includegraphics[width=0.45\textwidth]{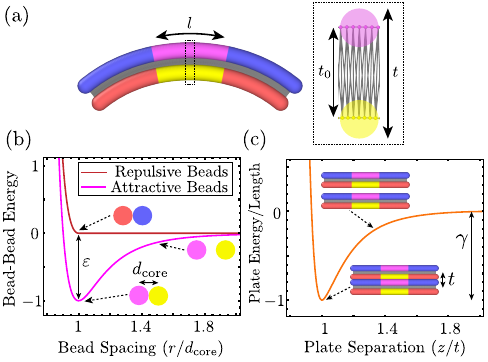}
    \caption{(a) Coarse-grained curvamers are modelled as a bead-spring network, with structural thickness $t_0$ between bead layers.  An attractive patch of length $l=w/3$ keeps particles aligned while stacking and prevents lateral sliding.  (b)  Bead-bead interactions between particles are composed of an attractive Lennard-Jones potential (shown for yellow-magenta beads) that is localized to the beads in the middle third of the particle and a purely repulsive Weeks-Chandler-Anderson potential for all other bead-bead interactions (shown for red-blue beads) .  All beads have a hard-core diameter of $d_{\rm core}$ which fixes their equilibrium separation distance. (c)  The interaction energy per unit length between two flat plates as a function of the center-center separation distance is obtained by summing over all the bead-bead interactions.  } 
    \label{fig:computermodel}
\end{figure}

For comparison to the continuum model of stacks, we perform numerical coarse-grained calculations of curvamer stacks with finite ranges of attraction based on a bead-spring model of discrete curvamers, following the design of Ref. \cite{conformalcurvamers}.  The positions of the evenly spaced beads (150 per layer) are placed along the top ($+$) and bottom ($-$) edges of the particle with radii $r_{\pm} = r_0 \pm t_0/2$, where the particle has structural thickness $t_0$.  This forms a trapezoidal truss network (see Fig. \ref{fig:computermodel}a) of three different types of springs (denoted $k_h$, $k_v$, and $k_c$, for horizontal, vertical and cross, respectively), whose rest lengths correspond with the particle having a preferred radius of curvature $r_0$ at its mid-line, and spring constants chosen such that effective elastic properties (i.e. bend to stretch modulus ratio) of the particle behave like an elastic shell of Poisson ratio of $\nu = 0.3$ (see Appendix B of \cite{conformalcurvamers}).  

As observed in Ref. \cite{conformalcurvamers}, some regimes of curvamer assembly become unstable to complex patterns of lateral particle sliding that have the effect of ``defocusing" curvature, and thus suppressing the range of self-limitation.  To suppress this mode and favor alignment of curvamer particles on the same axis, we introduce two types of bead-bead interactions (Fig. \ref{fig:computermodel}b).  The first is an attractive interaction using a shifted Lennard-Jones (LJ) potential
\begin{equation}
\label{equation:beadlj}
    U_{\text{LJ}}(r) = 4\, \varepsilon \left[ \left(\frac{\sigma}{r - \Delta r}\right)^{12} - \left(\frac{\sigma}{r - \Delta r}\right)^6 \right]\text{,}
\end{equation}
where $\varepsilon$ is the interaction strength, $\sigma$ is the range of the attractive well, and $\Delta r$ is the shift parameter that controls the equilibrium separation distance (hard core diameter) between attractive beads.  We define $d_{\text{core}}$ to be the minimal energy separation of the shifted LJ potential which we hold constant for variable interaction range via the relation $\Delta r = d_{\text{core}}-2^{1/6} \sigma$.  This attractive interaction is localized to the beads in middle of the particle, forming a ``sticky patch'' of length $l = w/3$ to prevent any lateral sliding between curvamer particles.  The remaining beads interact repulsively with a Weeks-Chandler-Anderson (WCA) potential 
\begin{equation}
\label{equation:beadwca}
    U_{\text{WCA}}(r) = \begin{cases}
        U_{\text{LJ}}(r) + \varepsilon\text{,} & r\leq d_{\text{core}} \\
        0\text{,} & r > d_{\text{core}}
    \end{cases}
\end{equation}
which smoothly goes to zero at $r = d_{\text{core}}$, and otherwise shares the same variables $\varepsilon$ and $\sigma$ as the attractive potential.  Additionally, bead-bead interactions within the same curvamer particle are turned off.  This choice of potentials means that when two particles interact and maintain perfect contact between their surfaces, only the beads in the sticky patch will contribute to the total cohesive energy, and repulsive flanks of the particle contribute only when they are overlapping.  The interaction strength $\varepsilon$ is chosen such that when two flat curvamers are separated by their effective thickness $t \approx t_0 + d_{\text{core}}$, the total cohesive energy is $-\gamma w$ which is held at a constant value, while the the dimensionless ratio of cohesion to intra-particle stiffness, $S$, is varied by decreasing the intra-particle spring constants.     

To map the coarse-grained parameters to their theoretical counterparts, we assume that $\gamma$ is roughly independent of particle curvature and calculate the particle bending modulus $B$ from the energy of a curvamer with preferred curvature $\kappa_0$ being placed in a completely flattened state, allowing us to compute the reduced cohesion $S$. By calculating the interaction energies of two flattened curvamers at various separation distances around the particle thickness $t$ (i.e. their equilibrium spacing), and fitting a parabola to the minimum of the interaction well,  we compute $\gamma''$ and thus map to the reduced gap stiffness $G$ (see Appendix \ref{appendix:cohesion} for more details).  Notably, we expect $\gamma'' \sim \sigma^{-2}$ so that gap stiffness is largely controlled by the {\it range} of the attractive well.  Since the particle geometry and cohesion energy is fixed, we ultimately vary $S$ by changing the spring constant in the coarse-grained model and vary $G$ through the bead-bead interaction range $\sigma$ and the spring constant.  Additionally, we find the relation between the effective interaction range between flat curvamer plates and the range of the LJ attractive well to be $\sigma_{\rm eff}\simeq 0.11\, \sigma$.

To calculate the equilibrium state and energy of a finite-sized stack, $N$ coarse-grained curvamers are uniformly stacked vertically with their centers aligned and curvatures decreasing according to curvature focusing so that they initially have perfect contact.  We then perform an energy minimization of the bead positions using a conjugate gradient algorithm with \texttt{LAMMPS} \cite{LAMMPS} to obtain the structure configuration that corresponds to a minimum in the energetic landscape.     

\section{Mechanically equilibrated structures and the gap opening transition}
\label{section:structures}

\begin{figure*}
    \centering
    \includegraphics[width=0.9\textwidth]{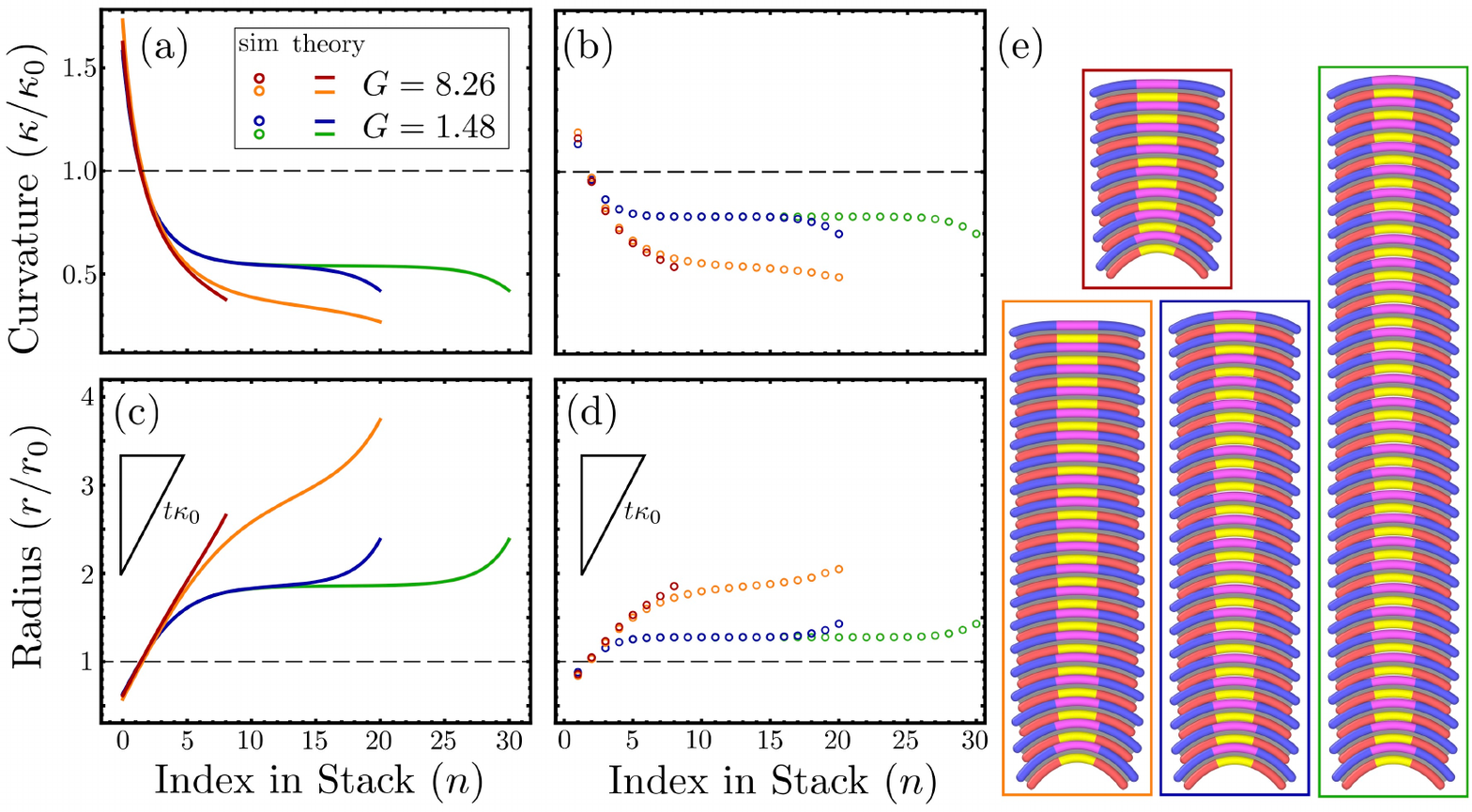}
    \caption{(a)-(b)  The curvature profiles of continuum model (lines) and coarse-grained (dots) curvamer stacks for three different stack sizes and two dimensionless ratios of gap stiffness to particle stiffness $G$.  The curvature at the ends of the stack are equally distant from the preferred curvature (dashed line).  (c)-(d)  The radius of curvature profile of curvamer stacks.  A slope of $t\kappa_0$ represents curvature focused stacking (gap-closed).  All stacks form a curvature focusing boundary layer at their ends.  Larger stacks and those with smaller $G$ see a clear deviation away from curvature focusing indicating the formation of gaps between particles with the middle section of the stack approaching nearly uniform shape.  (e)  Visualizations of the coarse-grained stacks plotted in (a)-(d).}
    \label{fig:curvatureprofile}
\end{figure*}

As mentioned in Sec. \ref{section:continuummodel}, by looking at the form of the total energy density in Eq. (\ref{equation:econtinuum}), we see that the two bare (i.e. distortion-free) cohesive terms do not depend on particle curvature, so solving Eqs. (\ref{eq: ELeq}) and (\ref{eq: BCs}) for the mechanically equilibrium curvature profile $\tilde{\kappa}_*(h)$ is controlled only by the scaled height $H$ of the stack and the dimensionless ratio of gap to particle stiffness $G$.  We start by considering two heuristic limiting behaviors as a function of gap stiffness.

First, for infinitely stiff gaps, $G \rightarrow \infty$, from Eq. (\ref{equation:eexcess}), it is straightforward to see that the excess energy is minimized when $\tilde{\kappa}'(h) = -\tilde{\kappa}^2(h)$ for all stacks, which is the curvature focusing condition of Eq. (\ref{equation:curvefocus}) in the continuum limit.  Thus, the particles stack with perfect contact and our theory reduces to that of conformal stacking of Ref. \cite{conformalcurvamers} when the inter-particle bonds are infinitely stiff and the excess energy is generated purely from particle shape deformations. The residual energy of this solution derives from the gradient of particle shape through the stack which grows monotonically with size,  i.e., for narrow curvature-focusing stacks ($H \ll 1$) it is straightforward to show the linear profile $\tilde{\kappa}_*(h) \approx (H/2-h)$.   In the opposite limit of infinitely compliant gaps, $G \rightarrow 0$, the $\epsilon_{\rm ex}[\kappa_*(h)]$ for all stack sizes favors uniformly undeformed curvatures, $\kappa(h) = \kappa_0$.  On these basic grounds we expect a transition in the equilibrium profiles from curvature focusing  to gap opened states as a function of decreasing $G$.


In Fig. \ref{fig:curvatureprofile}, we plot four mechanically-equilibrated profiles $\kappa(n)$, here as a function of unscaled parameters.  Comparing two different stack sizes with the same gap stiffness, we see that the smaller stack (red) has a nearly constant slope for its curvature radius profile (Fig. \ref{fig:curvatureprofile}c), which indicates it is close to the concentric stacking $r_n = r_- + nt$ condition required by curvature focusing.  In comparison,  for the larger stack (orange) with the same gap stiffness ($G=8.26$), curvature radii are curvature focusing only at the bottom and top, following a constant slope similar to the shorter stack, but with interior middle particles that deviate away from this slope becoming more uniformly shaped.   This flattening of the interior curvamer shapes is more evident when the gap stiffness is further reduced ($G=1.48$) for this same larger stack size (blue).  In this case, the particles become nearly uniform and adopt a curvature even closer to the preferred value $\kappa(n) \approx \kappa_0$ over a large segment of the shape profile in the middle of the stack.  Once again the top and bottom of the stack feature a return to the constant slope of curvature focusing.  As discussed above, the condition of free boundaries of Eq. (\ref{eq: BCs}) requires stacking to maintain curvature focusing at the top and bottom of the stack.  Hence, for increasingly large stacks with sufficiently compliant gaps (i.e. low enough $G$) we see that this curvature-focusing region becomes essentially a boundary layer in equilibrium curvature profiles.  This can be seen, by considering in Fig. \ref{fig:curvatureprofile}c, an even larger stack (green) profile for the compliant gap $G=1.48$, which evidently attains the same constant curvature value $\kappa(n) \approx \kappa_0$ in the stack interior, but over a larger length than the shorter (blue) stack, and is flanked by approximately the same curvature focusing regions on its free boundaries.  Last we note from Fig. \ref{fig:curvatureprofile}a that all of the analytical solutions for $\kappa_*(n)$ satisfy Eq. (\ref{eq: overunder}), with the magnitude of {\it overbending} at the bottom of the stack equal to the magnitude of {\it underbending} at the top, i.e. $\kappa_- - \kappa_0=-(\kappa_+ - \kappa_0)$.


In Fig. \ref{fig:curvatureprofile}b and d, we see that the coarse-grained simulations are in good agreement with the continuum theory predictions with regards to these basic features, although there are also some quantitative differences that arise from two aspects of the continuum to discrete comparison.   First, the boundary conditions in the continuum model strictly speaking are formulated in terms of  ``virtual particles'' at the edges of the stack (e.g. $n=0$), a position which is not actually sampled in the real discrete stacks.  Second, and related to this, since the reduced stack size is $h = n t\kappa_0$, the smallest resolution achievable in the coarse-grained simulations of stack size is that of one curvamer, or $\Delta h = t\kappa_0$.  Hence, if the boundary layer is small compared to this size scale (i.e. if the number of discrete particles in the boundary zone is not large), then the shape gradients in these regions will differ more significantly between the continuum and discrete models.  In both cases, these limitations can be reduced and effects be better seen by decreasing $t\kappa_0$ of the particles, at the expense of having to include more particles in the assembly to achieve the same reduced stack size $H = Nt\kappa_0$.  Nevertheless, the agreement shown in Fig. \ref{fig:curvatureprofile} suggests that the solutions of continuum model capture the essential features of intra-particle relative to inter-particle modes of distortion in frustrated stacks and their dependence on structural and elastic parameters.      

\begin{figure*}
    \centering
    \includegraphics[width=\textwidth]{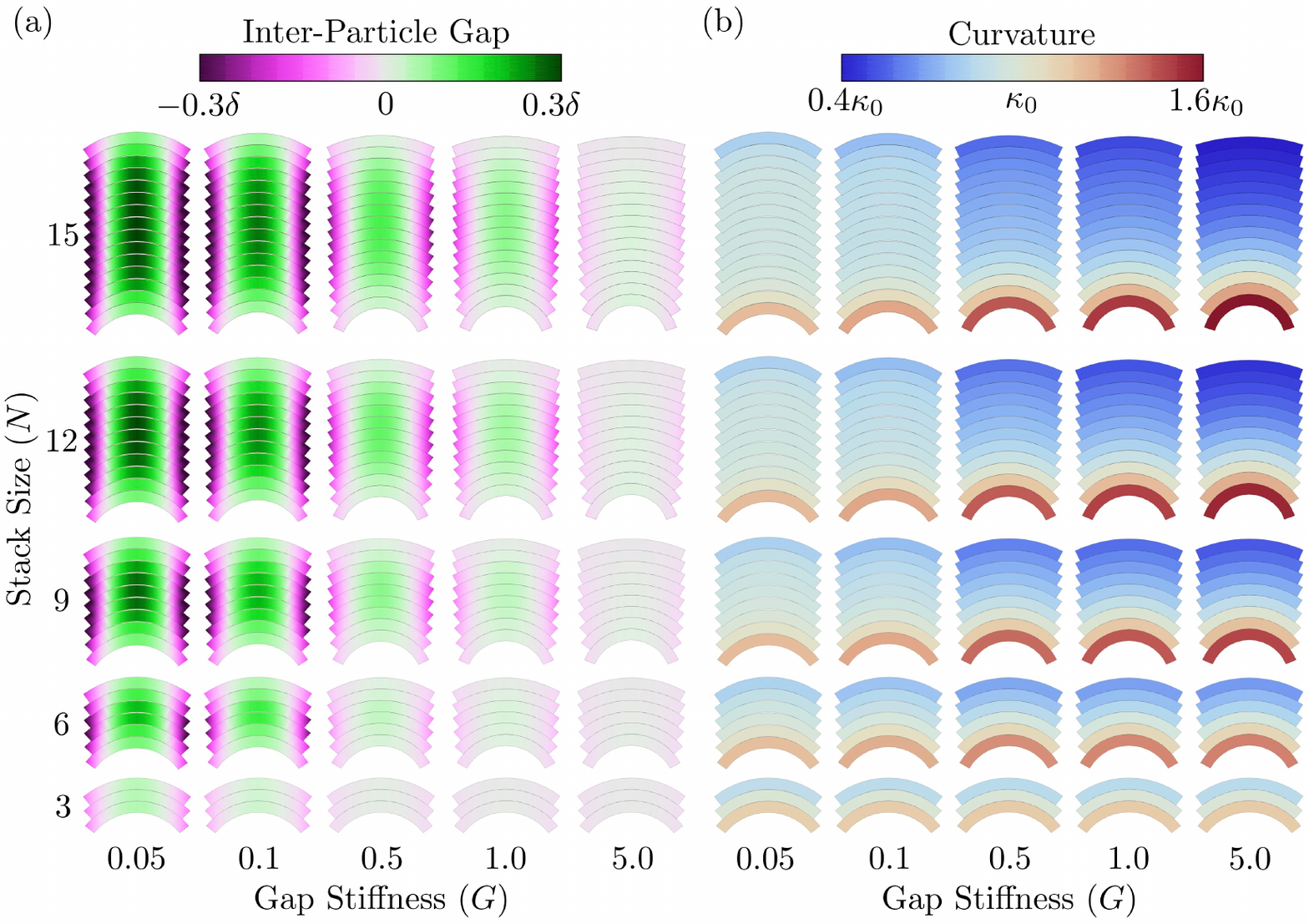}
    \caption{Mechanically-equilibrated stacks of different sizes $N$ (vertical axis) and gap to particle stiffness ratios $G$ (horizontal axis) colored by inter-particle gap distance in (a) and particle curvature in (b).  Stacks see their inter-particle gaps stretch open after a certain stack size $H_{\rm gap}(G)$ which increases with the gap stiffness $G$.  Stacks with high gap strain (gap opened) see low curvature strain (uniform shaped), while those with high curvature strain (curvature focused) see low gap strain (gap closed).}
    \label{fig:stackmatrix}
\end{figure*}

These example cases show the shape profile deviates from conformal, curvature-focusing stacking when $G$ is sufficiently low and when stack size is sufficiently large.  In these cases of non-conformal stacks, we therefore expect a variable degree of inter-particle strain.  The patterns of inter- and intra-particle strain, as well as the overall dependence on $G$ and $N$, in equilbrium stacks is illustrated in Fig. \ref{fig:stackmatrix}.  For lower $G$ and large $N$, Fig. \ref{fig:stackmatrix}a shows that stacks develop the largest magnitude of gap strains, but in general these vary both along the stacks, as well as along bound curvamers themselves.  This is because when curvamers are not curvature focusing, according to Eq. (\ref{eq: gapstrain}) the cohesive gap between them varies quadratically with lateral position, yet the mean gap between a particle pair $\Delta z_*$ adjusts so that the {\it net force} is zero.  Therefore, for non-focusing geometries, we observe a characteristic gap strain that is tensile (with gaps pulled open) at the center of particles and compressive (with curvamers pushing into one another) in their outer flanks.  Notably, in these same regimes where gap strain is highest (low $G$ and large $H$), Fig. \ref{fig:stackmatrix}b shows that curvature is most uniform and tends towards the preferred particle shape $\kappa(n) \to \kappa_0$, with the exception of the curvature-focusing boundary layer at the ends.  In contrast, as gap stiffness increases or stack size decreases, we observe the magnitude of gap strain visibly decrease, and a more obvious gradient in particle shape develops through the particle stack.  Taken together, Fig. \ref{fig:stackmatrix} shows that the mode of elastic distortion that absorbs the predominant effect of frustration in a curvamer stack, whether that be intra-particle bending or inter-particle gap strain, exhibits a complex inter-dependence on stack size and relative gap to particle stiffness.

\begin{figure}
    \centering
    \includegraphics[width=0.48\textwidth]{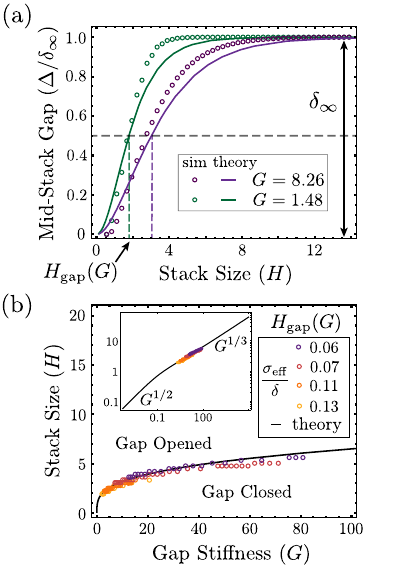}
    \caption{(a) The central gap between particles in the middle of a stack, $\Delta$, grows with stack size until becoming uniformly spaced ($\delta_{\infty}(G)$) for infinitely large stack sizes.  The gap opening transition size $H_{\rm gap}$ is defined to be when $\Delta = 0.5 \,\delta_{\infty}$.  (b) Continuum and coarse-grained model calculations show $H_{\rm gap}$ increases with gap to particle stiffness ratio $G$.}
    \label{fig:gapopening}
    
\end{figure}

In general, we can characterize this dependence as a {\it gap opening} transition at a characteristic stack size $H_{\rm gap}(G)$, from curvature-focusing/gap-closed stacks for $H \ll H_{\rm gap}(G)$ to uniform-shape/gap-opening profiles for $H \gg H_{\rm gap}(G)$ .  We characterize this transition in shape profile as shown in Fig. \ref{fig:gapopening}a, where we plot the central gap in the middle of the stack for fixed dimensionless gap stiffness and increasing stack sizes $H$, in general showing gap strain increasing from zero up to to a maximal gap size $\delta_\infty$ as $H \to \infty$.  We define $H_{\rm gap}(G)$ as the stack size at which the central mid-stack gap is half of $\delta_\infty$.  In Fig. \ref{fig:gapopening}b, we plot $H_{\rm gap}(G)$ from continuum theory, as well as a comparison to a range of simulated stacks, as a function of dimensionless gap stiffness, showing that this characteristic size increases monotonically with $G$ consistent with two power-law regimes:  $H_{\rm gap}(G)\sim G^{1/2}$ for $G \ll 1$ and $H_{\rm gap}(G)\sim G^{1/3}$ for $G \gg 1$.  This shows that conformal stacking is favorable for sufficiently small stacks for any gap stiffness, but also that this packing gives way to one that favors gap strain between uniformly shaped particles at a stack size that becomes smaller as inter-particle cohesion becomes relatively more complaint than particle shape.  

An understanding of when this transition occurs can be found via the following scaling argument that compares the energetics of two competing morphologies: gap-opened, uniform stacks at large size and conformal, curvature-focusing stacks at small size.  In the former case, the nature of infinitely large, uniform stacks is determined by considering the optimal constant curvature $\tilde{\kappa}_\infty$ in Eq. (\ref{equation:eexcess}) for the case $\tilde{\kappa}'=0$, which is governed by the roots of the cubic equation
\begin{equation}
\label{equation:kinfcubic}
    \tilde{\kappa}_\infty +2 G \tilde{\kappa}_\infty^3 =1 .
\end{equation}
In the case of small $G$, stiff particles with flexible bonds mean that uniform stacks  will have particles which only slightly flatten from their preferred shape $\tilde{\kappa} \simeq 1-2 G$, so that the non-conformal excess energy density scales as $\epsilon_{\text{ex}}^{\infty} \sim G$.  Meanwhile, for small conformal stacks, curvature changes linearly around a central particle with curvature $\kappa_0$, and the conformal excess energy then scales as $\epsilon_{\text{ex}} \sim H^2$.  The gap opening transition size, where these scalings cross over, will then go as $H_{\text{gap}} \sim \sqrt{G}$.  In the case of large $G$, it can be shown that flexible particles with stiff bonds will be nearly flat in infinite assemblies with uniform curvature $\tilde{\kappa}_{\infty} \sim G^{-1/3}$ implying large stacks approach absolute flattening for stiff gaps leading to a residual cost, $\epsilon_{\text{ex}}^{\infty} \sim \frac{1}{2} - G^{-1/3}$.  For large conformal stacks, curvature focusing implies $\epsilon_{\text{ex}} \sim \frac{1}{2} - \frac{\ln(H)}{H}$~\cite{conformalcurvamers} and so we find the gap opening transition will go as $G \sim (\ln(H_{\text{gap}})/H_{\text{gap}})^{-3}$, or up to a logarithmic correction, $H_{\text{gap}} \sim G^{1/3}$.  Notably, the stack size of this gap-opening transition diverges as $G \to \infty$, consistent with an asymptotic approach to the strictly conformal limit at all scales.

\begin{figure}
    \centering
    \includegraphics[width=0.48\textwidth]{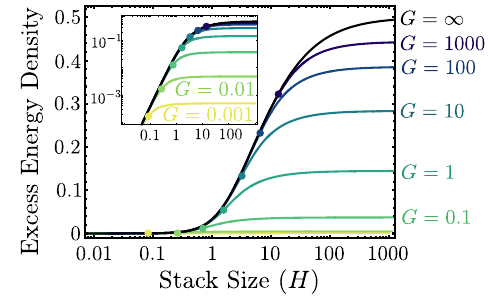}
    \caption{The excess energy density which penalizes curvature changes and gap stretching as a function of stack size for different gap to particle stiffness ratios $G$.  Stacks with finite values of $G$ are gap closed and curvature focusing for small sizes, then become gap opened at $H_{\rm gap}$ (dots) asymptotically approaching a finite energy density for infinitely large stacks.  The black curve represents the gap-closed, curvature focusing theory of Ref. \cite{conformalcurvamers}.}
    \label{fig:excessenergy}
\end{figure}

This gap-opening transition and the scaling picture that describes it are also reflected in the ultimate excess energy dependence on stack size predicted by the continuum model, plotted in Fig. \ref{fig:excessenergy}.  For all $G$, the limit of small stacks ($H \ll H_{\rm gap}$), exhibits conformal stacking and therefore a monotonically increasing cost of curvature focusing, e.g. $\epsilon_{\rm ex} (H \ll 1) \sim H^2$.  For large stacks  ($H \gg H_{\rm gap}$) excess energy accumulation plateaus due to a transition to a state which is predominantly gap-opened and uniform shape in the bulk of the stack, flanked by curvature-focusing boundary layers as illustrated in Fig. \ref{fig:curvatureprofile}.  As argued above, the energy cost of these infinite uniform stacks {\it decreases} with gap stiffness from a maximum of $\epsilon_{\text{ex}}^{\infty} (G \to \infty) \to 1/2$ for rigid gaps to a cost that ultimately vanishes with $G$ as $\epsilon_{\text{ex}}^{\infty} (G \ll 1) \sim G$ for compliant gaps.  Hence, the effect of increasing the flexibility of cohesive interactions is to reduce both the {\it size} range (in terms of stack size) and {\it energetic cost} of frustration accumulation in curvamer stacks.  In the following section, we analyze the thermodynamic consequences of this dependence of accumulation cost on relative stiffness of cohesion to particle shape deformation.

\section{Self-limiting vs unlimited assembly}
\label{section:sla}
In the canonical ensemble and for sufficiently saturated conditions (i.e. above the aggregation concentration threshold), equilibrium assembly is determined by the aggregate size that minimizes the free energy per subunit of interactions within the aggregate structure \cite{grason-slareview-2021, israelachvili_intermolecular_1992}.  
For curvamer stacks, the self-limiting assembly size, $H_*$, is defined to be the size for which the per subunit assembly energy of Eq. (\ref{equation:econtinuum}), is a minimum, which includes the per subunit ``surface'' cost $S/H$  which competes with $\epsilon_{\rm ex} (H)$ to set the optimal size.  The bulk term $\epsilon_0 = S/t\kappa_0$ only shifts $\epsilon (H)$ by a constant and therefore has no effect on the self-limiting size.  If no finite minimum occurs, then the assembly will be unlimited, which in the case of 1D stacking assembly strictly speaking corresponds to exponentially distributed lengths that grow with total concentration.    In this section, we will look at self-limiting stacks of curvamers and the thermodynamic regimes of self-limiting or unlimited assembly.  We first consider, in sec. \ref{section:sla-g}, self-limitation in the thermodynamic ensemble of {\it fixed ratio of gap to particle stiffness} $G = \frac{\gamma'' w^4 t^2 \kappa_0^2 }{ 720 B}$, and show that the self-limiting size is effectively tuned with the reduced cohesion $S = \frac{\gamma t}{B \kappa_0}$, up to a maximal value $S_{\rm max}(G)$ beyond which the assembly is driven to unlimited size.  We then consider and analyze, in sec. \ref{section:sla-k}, the more experimentally relevant ensemble, where effective interaction range, $\sigma_{\rm eff} = \sqrt{\gamma/\gamma''}$, is held constant, in which case the relevant fixed dimensional measure of gap compliance is defined relative to cohesive strength as we detail below.  

\subsection{Fixed ratio gap to particle stiffness}
\label{section:sla-g}

\begin{figure}
    \centering
    \includegraphics[width=0.48\textwidth]{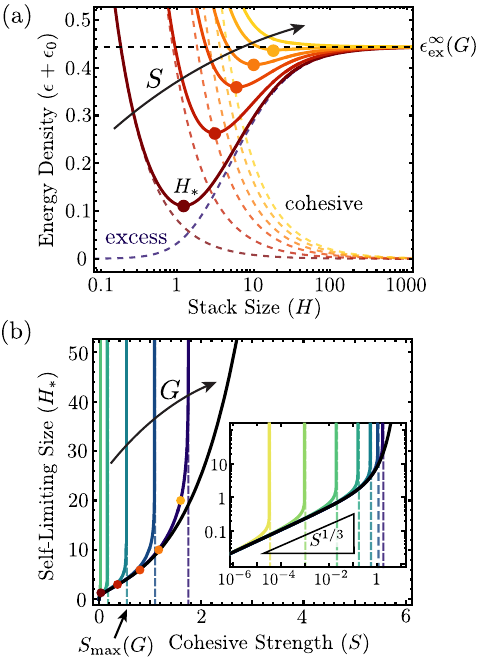}
    \caption{(a) The total energy density of the continuum model for $G=1000$. The minimum of the energy density curve is the self-limiting stack size $H_*$ (dots), which increases with cohesive strength to particle stiffness ratio $S$ and diverges as $S$ approaches a maximal cohesive strength $S_{\rm max}$.  Above this maximal value (yellow curve), assembly is unlimited.  (b)  Self-limitation occurs within a finite range of cohesive strength $S$ which increases with the gap stiffness $G$.  The curve colors correspond to the same values of $G$ shown in Fig. \ref{fig:excessenergy}, with the black curve representing gap-closed, curvature focusing assembly.}
    \label{fig:energydensity}
\end{figure}

 We first consider self-limitation in the case of fixing the ratio of inter-particle to intra-particle stiffness $G = \frac{\gamma'' w^4 t^2 \kappa_0^2 }{ 720 B}$ constant and increasing the dimensionless ratio of cohesion to flattening energy $S$.  In this case, the excess energy density takes the form of the fixed-$G$ solutions described in the previous section (see Fig. \ref{fig:excessenergy}), so if and where a minimum occurs will depend on the strength of the cohesive boundary penalty $S/H$ which generically favors larger stack sizes and implies a self-limiting size $H^*(S,G)$ that grows with $S$.  As an example, we consider a case of fixed $G=10^3$ in Fig. \ref{fig:energydensity}a for an increasing range of $S$. Whether a minimum in $\epsilon(H)$ occurs at finite $H$ depends on how fast the excess energy density grows with size and is outlined for a general $d$-dimensional frustrated assembly in Ref. \cite{grason-slareview-2021}.  It is straightforward to show (Appendix \ref{appendix:selflimiting-appendix}) that self-limiting states are described by the equation of state relating optimal size $H_*$ to the reduced cohesion
\begin{equation}
    \label{equation:ssla}
    S (H_*)= H_*^2\frac{d\epsilon_{\rm ex}}{dH} \bigg|_{H=H_*} \text{.}
\end{equation}
For the limit of very small (i.e. curvature focusing) stacks we expect $\epsilon_{\rm ex}\sim H^2$, from which it is straightforward to show the power-law growth of self-limiting size with cohesion
\begin{equation}
\label{eq:small}
 H_*(S \to 0) \sim S^{1/3} .
\end{equation}
As illustrated for the case in Fig. \ref{fig:energydensity}a, however, for sufficiently large $S$ the self-limited size increases and eventually exceeds $H_{\rm gap}$, meaning that the relevant stack profile is approaching the uniform curvature state for which excess energy plateaus, i.e $\epsilon_{\rm ex} (H \gg H_{\rm gap}) \simeq \epsilon_{\rm ex}^{\infty} (G)$.  Since $\frac{d\epsilon_{\rm ex}}{dH} \approx 0$ for $H \gg H_{\rm gap}$, the accumulation of elastic energy with size is not sufficient to balance the cohesive drive to increase stack size at these scales, and hence, assembly is no longer self-limiting for this range of cohesion.   This behavior can be seen clearly in Fig. \ref{fig:energydensity}a, where for large $S$ the energy density monotonically decreases approaching the bulk energy $\epsilon_{\rm ex}^{\infty} (G)$ as the structure grows infinitely large (i.e. $H_* \to \infty$).  A simple estimate for this maximal cohesion $S_{\text{max}}(G)$ for self-limitation is given by $ H_*\sim S^{1/3} \approx H_{\rm gap} (G)$. For small $G$, this implies $S_{\text{max}}(G \ll1) \sim G^{3/2}$, illustrating that the cohesive range for self-limiting assembly grows with gap stiffness, consistent with an analogous argument for stiff gaps which suggests that $S_{\text{max}}(G \gg 1) \sim \ln G $.  Exact calculations of $S_{\rm max}(G)$ can be seen in Fig. \ref{fig:Smax} of Appendix \ref{appendix:selflimiting-appendix-k}.


A careful analysis (see Appendix \ref{appendix:selflimiting-appendix} for detailed calculation) shows that the  self-limiting  size diverges continuously at this maximal cohesion $S\rightarrow S_{\text{max}}(G)$ , i.e. there is a second-order transition which occurs at $S_{\text{max}}(G)$ between self-limited to unlimited assembly.  Equations of state $H_*(S)$ versus $S$ are shown for a series of $G$ values in Fig. \ref{fig:energydensity}b.  This is in contrast to the conformal theory where the self-limiting size is finite for any finite $S$ and strictly only diverges at $S\rightarrow \infty$ (black line in Fig. \ref{fig:energydensity}b).  From Appendix \ref{appendix:selflimiting-appendix}, we find a logarithmic divergence of self-limiting stack size at the transition $H_* (S \to S_{\text{max}}) \sim - \ln \left(S_{\text{max}}-S\right)$.  

\subsection{Fixed attraction range: phase diagram of self-limiting behavior}
\label{section:sla-k}

The analysis of the prior section considers self-limiting assembly of curvamer stacks under conditions of two indepedent dimensionless variables: the ratio of cohesive strength to particle stiffness, $S = \frac{\gamma t}{B \kappa_0}$, and the ratio of cohesive stiffness to particle stiffness, $G = \frac{\gamma'' w^4 t^2 \kappa_0^2 }{ 720 B}$. However, for physical models of colloid curvamer assembly, we argue that it more useful to consider a different ensemble, one in which the two parameters that describe the particle interactions ($\gamma$ and $\gamma''$, or the respective {\it depth} and {\it stiffness} of binding) are combined into a single dimensionless parameter that is independent of the elastic properties of the particles.  As discussed below, this is motivated by the fact that for colloidal forces that can be used to drive curvamer binding (e.g. depletion or surface-functionalized DNA linkers), the {\it range} of attractive interactions is held fixed, even while the strength of attractions is variable.  In such cases, $\gamma$ and $\gamma''$ do not vary independently and as illustrated in Fig. \ref{fig:harmonicapprox}c, the their ratio defines a characteristic lengthscale of binding, $\sigma_{\rm eff} = \sqrt{\gamma/\gamma''}$ a generic measure of the interaction range.  To consider the case of fixed interaction we therefore introduce a new dimensionless measure of gap stiffness
\begin{equation}
\label{equation:reducedK}
K \equiv G/S =  \frac{w^4 t \kappa_0^3}{ 720}\sigma_{\rm eff}^{-2} \propto \frac{1}{t\kappa_0}\left( \frac{\delta}{\sigma_{\rm eff}}\right)^{2} ,
\end{equation} 
which quantifies the ratio of surface energy in a stack of thickness $r_0$ to the cost of cohesive strain induced by the ``natural gap'' $\delta \simeq \kappa_0^2 w^2 t/8$  between particles with their ideal shape.  

\begin{figure}
    \centering
    \includegraphics[width=0.48\textwidth]{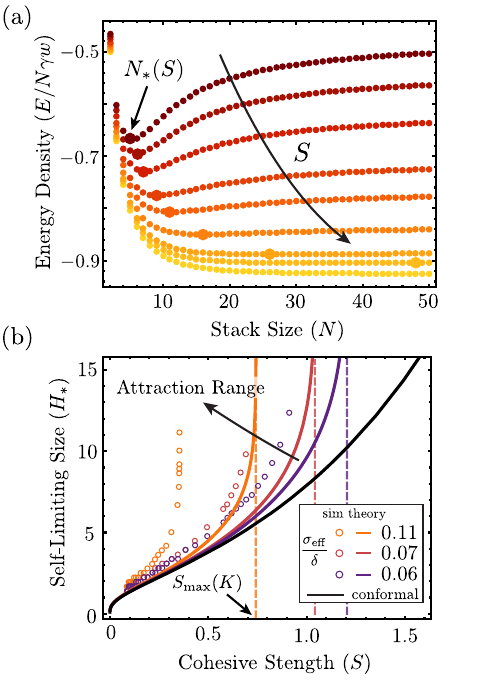}
    \caption{(a) Coarse-grained simulation results for particles with fixed effective interaction range ($\sigma_{\rm eff}/\delta$ = 0.07, K = 82.46).  Increasing the reduced cohesive strength $S$ increases the self-limiting stack size (larger circles) until a maximal cohesion, $S_{\rm max}(K)$, where no self-limitation occurs (yellow curve).  A minimum is counted as self-limiting if it is less than the maximum stack size calculated (50).  (b) Comparison of self-limiting sizes with finite interaction range for the coarse-grained (open circles) and continuum models (solid lines).  Both show increasing self-limiting sizes with $S$ which deviate away from the curvature focusing, conformally contacting limit (black curve).  Stacks are self-limiting within a finite range of cohesive strength $S$ which decreases with interaction range.  Values of $\sigma_{\rm eff}/\delta$ correspond to $K = 35,\,82,\,131$ for orange, red, and purple, respectively.}
    \label{fig:computerSLA}
\end{figure}

\begin{figure*}
    \centering
    \includegraphics[width=0.9\textwidth]{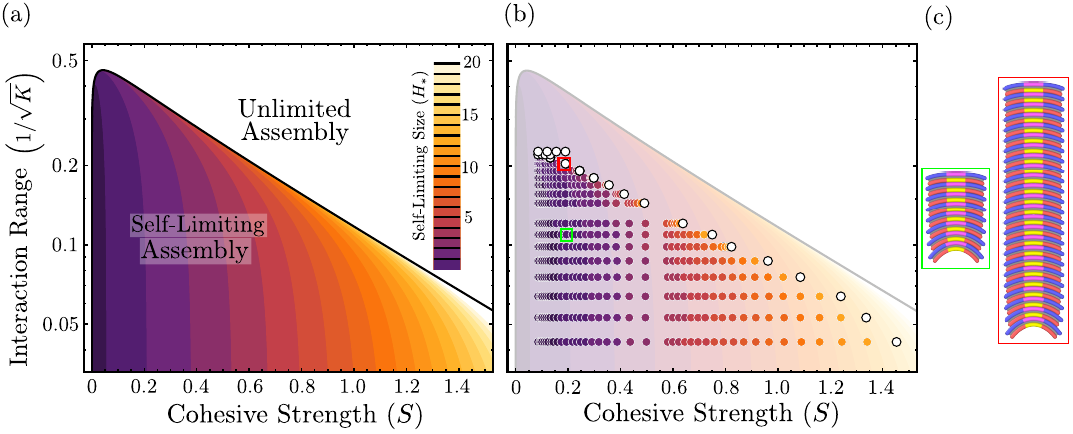}
    \caption{(a) Phase diagram of self-limiting assembly for curvamers with fixed interaction range.  Above a maximal reduced interaction range $1/\sqrt{K^*}\approx 0.447$, self-limiting behavior vanishes completely.  (b) Coarse-grained simulations qualitatively match the continuum model phase diagram and see vanishing self-limiting behavior above a critical reduced interaction range.  Filled circles represent self-limiting stacks.  Open white circles represent choices of $K$ and $S$ which did not see a minimum below the maximum stack size simulated (50) and are considered unlimited assembly.  (c) The green bordered stack is the self-limiting state for $K = 82$ and $S = 0.19$. For a longer interaction range at the same $S$, the self-limiting stack size is larger (red square, $K = 24$).}
    \label{fig:phasediagram}
\end{figure*}

It is straightforward to recast the thermodynamics of stack assembly in terms of fixed $S$ and $K$ and solve for the self-limiting stack size $H_*(S,K)$ (see Appendix \ref{appendix:selflimiting-appendix-k}).  
In Fig. \ref{fig:computerSLA}a, the total energy density curves from the coarse-grained curvamer simulations are plotted for one choice of interaction range corresponding to fixed value of $K=82$ and a sequence of increasing $S$ (here controlled by the ratio of intra-particle spring stiffness $k$ to LJ attraction strength).  In  simulations, we simulate stacks up to $N=50$ particles in size, and characterize the minima as self-limiting if $N_* <50$.  Fig. \ref{fig:computerSLA}b shows plots of (scaled) self-limiting stack sizes $H_*$ as a function of the reduced cohesion $S$ for different interaction ranges, corresponding to distinct fixed $K$ values.  Simulation and theory calculations are in qualitative agreement with the fixed-$G$ behavior shown in Fig. \ref{fig:energydensity}b and capture the self-limiting curves pulling away from the curvature focusing, conformal limit, with the point of deviation occurring at smaller sizes with increasing attraction range.  Although $G=KS$ is not constant in this sequence, for fixed $K$ it is still the case that the self-limiting stack size eventually reaches and exceeds the gap-opening size at an upper limit of cohesion.  Beyond this size range, we again find that self-limiting stack sizes diverge continuously at some maximal cohesion $S_{\rm max}(K)$.  We can find this maximal cohesion at fixed $K$ from our previous solution for $S_{\rm max} (G)$ via the solution to
\begin{equation}
    S_{\rm max}(K\cdot S) = S \text{.}
\end{equation}
For $K \geq 5.0$ we find that this relation has two solutions corresponding $S_{\rm max}(K)$ and $S_{\rm min}(K)$.  The minimal value corresponds to a low-$S$ regime where $G\to0$ so that gaps open, and assembly becomes unlimited, essentially at all stack sizes.  In practice, this $S<S_{\rm min}(K)$ regime corresponds to a narrow, if not completely negligible, region of the parameter space for colloidal curvamers.  For the upper limit to cohesion, as $S \to S_{\rm max}(K)$ we find the same power-law divergence of stack size as the case of fixed $G$.  Fig. \ref{fig:computerSLA}B shows that this maximal cohesive range decreases with increased range of interactions, both in the continuum theory (via decreasing $K$) as well as the discrete curvamer simulation model (via increasing $\sigma$).  


In Fig. \ref{fig:phasediagram}a, we plot a phase diagram of self-limiting behavior in terms of reduced interaction range $K^{-1/2} \propto \sigma_{\rm eff}$ and reduced cohesion $S$.  The continuum model predicts a transition line (shown in black) that separates self-limiting assembly (below) from unlimited assembly (above) defined by the parameterized curve $\left( S_{\rm max}(G), K = G/S_{\rm max}(G) \right)$.  Critically, there exists a maximum to this transition line at  $K^{-1/2} \approx 0.45$, predicting that above a maximal interaction range self-limiting assembly is not possible for any $S$.  
Below this critical interaction range, we see that the range of self-limiting parameter space is limited, but widens to arbitrarily large cohesive range as interaction range goes to zero (i.e. $K \to \infty$).  The maximal value of reduced interaction range $K^{-1/2}$ falls to zero exponentially with increasing $S$.   

In Fig. \ref{fig:phasediagram}b, we show coarse-grained simulation results for fixed curvamer dimensions but variable cohesive strength and interaction range.  Notably, simulation results are shown as filled color circles for the cases where a minimum in the energy density for $N_* < 50$ was found.  Parameters sampled at larger interaction range and cohesive strength where no such minimum was found are shown as open white circles.  Simulation results  are in good qualitative agreement with the continuum model predictions (transparent background) and notably also show an upper limit to self-limitation in the phase diagram, though at a slightly smaller value of the mapped value of reduced interaction range.  Additionally, we find that the upper cohesive range for self-limitation extends somewhat below the predicted value of $S_{\rm max} (K)$ from the continuum model, but nevertheless exhibits a similar increase in $S_{\rm max} (K)$ as $K^{-1/2}$ is reduced towards zero.

Fig. \ref{fig:phasediagram}b shows that there are quantitative differences between the predictions of the continuum model and the simulated stacks of coarse-grained curvamers. Most obvious is the depression of the range of self-limitation in terms of reduced interaction range $K^{-1/2}$.  For the coarse-grained, discrete curvamers, we find an upper limit of $K^{-1/2}\approx 0.22$ which is roughly half of the value predicted by the continuum theory.  While there are several non-linearities neglected in the continuum model that likely limit its accuracy, we expect the principle discrepancy derives from the harmonic approximation of the inter-curvamer surface potential.  Specifically, as illustrated in Appendix G, the true plate-to-plate potential is reasonably described by a quadratic approximation only very close to its minimum, while sampled assemblies experience strains far outside of this quadratic regime, especially near the self-limiting/unlimited boundary.  Notwithstanding the limitations of our minimal description of cohesive strain,  the continuum theory captures the non-trivial features of the self-limiting phase diagram, notably the exponential dependence of maximal interaction range on cohesive strength, as well as the generic dependence of self-limiting stack size on cohesion.


\section{Discussion and Conclusion}
\label{section:discussion}

In this article, we have studied two theoretical approaches to model the stacking assembly of flexible, curved, colloidal curvamers, in particular to assess the role of relative compliance of inter-particle forces (cohesion) to intra-particle shape (curvature).  The central conclusion is that any measure of inter-particle compliance leads to a new mode of ``frustration escape'', in which assembly thermodynamics favors unlimited stacks of uniform shape.  This behavior is controlled by the dimensionless ratio of cohesive to particle-shape stiffness, $G$, and is characterized by a threshold (reduced) stack size $H_{\rm gap}(G)$.  For $H \ll H_{\rm gap}(G)$, inter-particle gaps are suppressed and curvature-focusing in the stack leads to elastic energy that accumulates superextensively, i.e. faster than linearly with stack size.  For $H \gg H_{\rm gap}(G)$, gaps open in the bulk of stacks facilitating a uniform elastic energy growth of stacks.  Since the role of the excess energy is to counteract the cohesive drive to assemble, the result of the inter-particle mode of elastic deformation is to reduce the level of frustration allowing for larger self-limiting assembly sizes.  Consequently, this also means that gap-opened assemblies have a smaller range of cohesive strength that allows for self-limitation, as stack sizes continuously diverge to unlimited assembly at the cohesive strength to intra-particle stiffness ratio $S_{max}(G)$ that increases with gap size.  We analyzed the effects of this for the case of {\it fixed interaction range}, which is characterized by a distinct dimensionless quantity $K = G/S \sim \frac{1}{t\kappa_0}\left( \frac{\delta}{\sigma_{\rm eff}}\right)^{2}$, inversely proportional to the square of cohesive range $\sigma_{\rm eff}$.  Notably, curvamer stacks are predicted to be self-limiting for cohesive strengths up to an upper threshold $S_{\rm max}(K)$ that increases with $K$.  Most significantly, we find that self-limitation only occurs below maximal effective interaction $\sigma_{\rm eff} ({\rm max}) \simeq 0.13\, \delta / \sqrt{t \kappa_0}$, where $\delta$ is the ``natural gap'' between two non-overlapping curvamers of the same preferred shape.  In effect, sufficiently short-ranged cohesion is required in order to transfer ``shape misfit'' from one particle to the next.  For highly compliant gaps, long-range interactions maintain sufficient cohesion without gradients in particle curvature that accumulate with size.   Given this, self-limiting assembly is only possible for a cohesive interaction range below the critical value, and more generally, the cohesive range of self-limitation grows as the interaction become shorter ranged and effectively stiffer.  Below we discuss experiment implications of this result in the context of distinct colloidal designs of curvamer particles.  


We consider three class of particle designs, (a) photo-lithographically fabricated polymeric microshells \cite{tanjeem-shapechanging-2022,kuenstler-curvedhydrogels-2020,jeon-trilayergels-2020,na-greyscaleshapes-2016}, (b) banana-shaped colloidal particles \cite{fernandezrico-bananas-2020,fernandezrico-bananavortex-2021,fernandezrico-bananadiffusion-2023} and (c) curved particles derived from DNA origami \cite{dietz-foldingdna-2009,Han-dnacurvature-2011}.  These examples differ primarily in terms of structural dimensions with lithographically defined shells and banana colloids having width and thickness ranges of $w \approx 5-15\, \mu\text{m}$ and $t \approx 100-500\text{ nm}$, whereas DNA origami based curved particles can be made at order of magnitude smaller size $w \approx 50-500 ~{\rm nm }$ and $t \approx 5-50 ~{\rm nm }$.  Knowing the approximate dimensions and curvature of a curvamer design, we can apply the results of Sec. \ref{section:sla-k} and solve for the maximal effective interaction $\sigma_{\rm eff}(\rm max)$ that permits self-limiting assembly.  Here, we calculate the maximal range by assuming the particles are in a highly curved state with a radius of curvature such that the particle width forms approximately 45\% the circumference of a circle.   In general, the maximal interaction range will decrease for flatter particles so these estimates should provide good upper bound estimates for the range of attraction necessary for self-limiting curvamer stacks.  In the case of polymeric shells (a) with approximate radius of curvature $r_0 = 2\, \mu\text{m}$, the maximum effective interaction range which allows for self-limitation is $\sigma_{\rm eff}(\rm max)\approx 47 \text { nm}$. Banana-shaped colloids (b) with radius of curvature $r_0 = 6\, \mu\text{m}$ would need interactions less than $\sigma_{\rm eff}(\rm max)\approx 180 \text { nm}$ long to be self-limiting.  DNA origami particles (c) with approximate radius of curvature $r_0 \approx 180\text{ nm}$, on the other hand would need cohesive interactions with ranges less than $\sigma_{\rm eff}(\rm max)\approx 12 \text { nm}$ to be self-limiting.  This range of interaction lengths suggests short-ranged attractions such as depletion and single stranded DNA oligmers as two possible candidates for attractive interactions between physical curvamers.  In particular, depletion induced attraction might be well suited to the lithographic polymer shell and banana colloid particle systems as the interaction range (estimated as the depletant diameter) can vary from $\sim 5 \text{ nm}$ for SDS micelles and up to $100 \text{ nm}$ and beyond for non-adsorbing polymeric depletants and hard spheres made of polystyrene or silica \cite{Lekkerkerker2011, Zhao-ColloidalSelfAssembly-2018}.  Similarly, single stranded DNA oligomers (ssDNA) are well suited to the DNA origami system as they can programmed into the surface of a particle and hybridize with a complementary set of ssDNA on a separate particle's surface to form a bond with an interaction range (estimated as the length of the ssDNA) of approximately $5-20\text{ nm}$ \cite{ Zhao-ColloidalSelfAssembly-2018, rogers-dna-2011, videbaekk-economical-2024}.

In addition to the design considerations raised by our curvamer model, the predicted limit on the interaction range for self-limiting behavior likely raises additional questions about kinetic constraints for reaching self-limiting equilibrium states.  At present, our coarse-grained curvamer model has been used to sample the energetic groundstates prepared by pre-aligned stacks in close contact.  And while there is prior evidence that these states are stable to some measure of thermal fluctuations~\cite{conformalcurvamers}, it remains to be understood how the shape of curvamers and the straining of inter- and intra-particle bonds in equilibrium stacks influences the time scales necessary to find these equilibrium states.  In particular, the requirement for sufficiently short-ranged interactions may place additional constraints for reaching self-limiting equilibrium under experimentally relevant conditions of assembly at fixed concentration and temperature initiated for randomly dispersed states.  In general, as the range of attractive interactions is lowered, the kinetic cross-sections for two particles to bind decreases considerably, slowing down even unfrustrated assembly~\cite{Cheng_SoftMatter_2012, Hagan_ModelViral_2014}.  Additionally, it is generically true that self-limiting stacks themselves will likely be able to bind into hierarchical ``superstacks'', due to the weak and imperfect attractions between misfitting ends.  The possibility of ``weak binding'' of self-limiting domains through defective bonds is a generic feature of many discrete particle models of frustrated assembly~\cite{Spivack-puzzlemers-2022, hall-wedges-2023}.  The effect of weak binding in 1D frustrated assembly (e.g. stacking) has recently been predicted to lead to a minimal temperature for stable self-limitation, below which finite-size assemblies condensed into effectively unlimited chains~\cite{wang-polybricks-2024}.  The stability of ``weak binding" between self-limiting stacks of curvamers, as well as the influence of interaction range on kinetically accessible states of self-limitation motivate the need for finite-temperature dynamical simulation studies of curvamer assembly.

We conclude by noting that the frustration mechanisms underlying self-limiting assembly of curvamers is shared by a broader range of physical systems.  Indeed, the source of frustration in curvamers is common to a broad class of liquid crystalline systems, such as bent-core mesogens, whose shapes favor bending of the nematic director, without splay~\cite{Jakli_2018}.  The resulting ``bend nematic'' states and their geometric frustration is well appreciated~\cite{Niv_SoftMatter_2018, Snir_PRE_2022, Selinger_AnnRevCondMatt_2022}, leading to bulk states where frustration is either resolved by shape-flattening or defect-mediated modulated states.  However, curvamer assembly is more precisely related to smectic phases with preferred layer curvature~\cite{sethna_smectic_1982}, as inter-curvamer attractions penalize distortion or local spacing, as opposed to local alignment, of neighbor curvamers.  The effect of frustration in the multi-layer stacking by preferred curvature has been previously studied~\cite{didonna_blue_2003}, particularly in the cases smectic liquid crystal layers with preferred curvature~\cite{Achard_EPJE_05, Hough_HelicalNanofilament_2009, Matsumoto_Nanofilaments_2009}, and more recently, in the stacking assembly of nanosheets~\cite{Jana_NanoPlatlets_2017, serafin-polyhedra-2021}, where it has been argued to give rise to finite-domain size selection via a similar competition between elastic and surface energies.  Interestingly, in these examples, the preferred shapes have {\it zero mean curvature} and {\it negative Gaussian curvature}, and take the form of locally twisted helicoids.  This raises the basic question about how the nature of frustration propagation in stacking assembly varies for particles with preferred non-zero Gaussian curvature.  Geometric constraints of uniform layer stacking~\cite{didonna_blue_2003} require variable Gaussian curvature in a stack where any layer has non-zero Gaussian curvature, distinct from the present case of cylindrical curvature which requires only gradients in mean curvature.  Because changes of Gaussian curvature require changes in the metric of the layer, variable Gaussian curvature in conformal stacks would therefore require additional costs associated with layer {\it stretching} and will likely reshape the nature of excess energy accumulation.  Unlike cylindrical shells it is impossible for saddle- or sphere-shaped shells to escape frustration by flattening without potentially large costs of intra-layer stretching.  Therefore, it remains to be explored whether curvamers with non-zero Gaussian curvatures instead exhibit wholly distinct modes of frustration escape and qualitatively different regimes of self-limiting behavior.

 \section*{Acknowledgments}
        The authors are grateful to M. Stevens, M. Minnis, M. Wang and N. Hackney for valuable discussions and input.  This work was supported by US National Science Foundation through award NSF DMR-2028885 and the Brandeis Center for Bioinspired Soft Materials, an NSF MRSEC, DMR-2011846.  
\appendix

\section{Calculating curvamer stacks in mechanical equilibrium}
\label{appendix:curvprofile}

The total energy density of a stacking assembly of curvamers depends on the shape profile of the particles in the stack, specifically for the excess energy term which determines how both particle shape and inter-particle gap deformations are to be penalized.  By optimizing the excess energy functional of Eq. (\ref{equation:eexcess}) with respect to the shape profile, we also minimize the stacking energy for a curvamer stack of scaled size $H$ with dimensionless ratios of cohesion to particle stiffness $S$, and cohesive stiffness to particles stiffness $G$.  To find this optimal (dimensionless) curvature profile $\tilde{\kappa}(h)$ as a function of scaled height in the stack $h$, we take the variation of the excess energy functional of Eq. (\ref{equation:eexcess}), re-written here as 
\begin{equation}
\label{equation:excessfuntional}
\epsilon_{\rm ex} [ \tilde{\kappa}(h)] = \frac{1}{H} \int_0^H \mathcal{L}\big(\tilde{\kappa}(h),\tilde{\kappa}'(h)\big) \,{\rm d}h\text{ ,} 
\end{equation}
with
\begin{equation}
    \label{equation:excesslagrangian}
    \mathcal{L}\big(\tilde{\kappa}(h),\tilde{\kappa}'(h)\big) = \frac{1}{2}(\tilde{\kappa}-1)^2 + \frac{G}{2}(\tilde{\kappa}'+\tilde{\kappa}^2)^2 \text{ ,}
\end{equation}
and find
\begin{multline}
    \label{equation:excesvariation}
    \delta \epsilon_{\rm ex} = \frac{1}{H} \int_0^H \left( \frac{\partial \mathcal{L}}{\partial \tilde{\kappa}} - \frac{\rm d}{{\rm d}h}\frac{\partial \mathcal{L}}{\partial \tilde{\kappa}'} \right) \delta \tilde{\kappa}(h)\,{\rm d}h \\ 
    + \frac{\partial \mathcal{L}}{\partial \tilde{\kappa}'}(H)\delta \tilde{\kappa}(H) - \frac{\partial \mathcal{L}}{\partial \tilde{\kappa}'}(0)\delta \tilde{\kappa}(0) \text{ .}
\end{multline}

We then set $\delta \epsilon_{\rm ex} = 0$ for all $\delta\tilde{\kappa}(h)$, and find three equations which much be satisfied.  The first we identify as the Euler-Lagrange equation 
\begin{equation}
    \label{equation:eom}
    \tilde{\kappa}''-2\tilde{\kappa}^3
-\frac{1}{G}(\tilde{\kappa}-1) = 0\text{ ,}
\end{equation}
that governs curvature profiles of energy optimizing stacks which necessarily are in mechanical equilibrium.  The second and third we identify as the free boundary conditions 
\begin{equation}
\label{equation:bcH}
\tilde{\kappa}'(H) = -\tilde{\kappa}^2(H)\text{ ,}
\end{equation}
\begin{equation}
\label{equation:bc0}
\tilde{\kappa}'(0) = -\tilde{\kappa}^2(0)\text{ ,}
\end{equation}
at the top ($H$) and bottom ($0$) ends of the stack.  We note that both Eqs. (\ref{equation:bcH}) and (\ref{equation:bcH}) are of the form of Eq. (\ref{equation:curvefocus}) in the continuum limit, which describes a curvature focused stacking configuration.  Thus, we see that the ends of the stacks must be curvature focusing with closed gaps between particles.  

Multiplying Eq. (\ref{equation:eom}) by $\tilde{\kappa}'$ and integrating with respect to $h$, we obtain the squared rate of change of curvature in the stack 
\begin{equation}\label{equation:firstintegral}
    (\tilde{\kappa}')^2
 = \tilde{\kappa}^4 + \frac{1}{G}(\tilde{\kappa}^2-2\tilde{\kappa})+C\text{ ,}
\end{equation}
in terms of a conserved quantity $C$.  Evaluation at the ends of the stack using Eqs. (\ref{equation:bcH}) and (\ref{equation:bc0}) reveals 
\begin{equation}
    \label{equation:conservedquantity}
    C = -\frac{1}{G}\left[ (\tilde{\kappa}(H)-1)^2 - 1 \right] = -\frac{1}{G}\left[ (\tilde{\kappa}(0)-1)^2 - 1 \right]\text{ ,}
\end{equation}
and we find
\begin{equation}
    \label{equation:overunder}
    \big| \tilde{\kappa}_-- 1\big|^2 = \big| \tilde{\kappa}_+- 1\big|^2\text{ ,}
\end{equation}
where we've defined $\tilde{\kappa}_- \equiv \tilde{\kappa}(0)$ and $\tilde{\kappa}_+ \equiv \tilde{\kappa}(H)$ to be the curvature at the bottom and top of the stack respectively.  This surprising result tells us that the degree of deviation away from the preferred shape at the bottom of that stack equals that of the top of the stack and relates the end curvatures in a simple way,
\begin{equation}
    \label{equation:topcurvatre}
    \tilde{\kappa}_+ = 2 - \tilde{\kappa}_- \text{ .}
\end{equation}
We now write the squared rate of change of curvature in the stack in terms of the bottom curvature $\tilde{\kappa}_-$, and find
\begin{equation}
    \label{equation:Tke}
    T(\tilde{\kappa},G,\tilde{\kappa}_-)= \left(\frac{{\rm d}\tilde{\kappa}}{{\rm d}h}\right)^2 = \tilde{\kappa}^4 + \frac{1}{G}\Big((\tilde{\kappa}-1)^2 - (\tilde{\kappa}_--1)^2 \Big) \text{.}
\end{equation}

\begin{figure}
    \centering
    \includegraphics[width=0.4\textwidth]{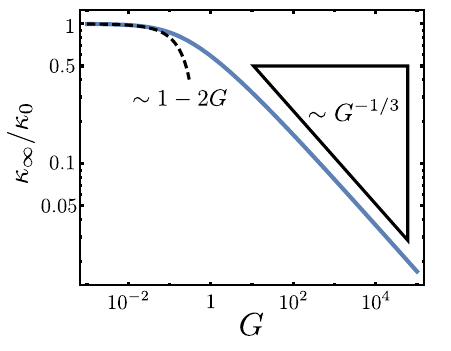}
    \caption{The (dimensionless) uniform curvature $\tilde{\kappa}_{\infty}$ of particles in an infinite stack as a function of reduced gap stiffness $G$.  Stacks with extremely compliant gaps ($G\to 0$) will have particles near their preferred shape ($\tilde{\kappa}_{\infty} \to 1$), while stacks with infinitely stiff gaps ($G\to \infty$) will have particles that are completely flattened ($\tilde{\kappa}_{\infty} \to 0$).}
    \label{fig:kinf}
\end{figure}

\begin{figure}
    \centering
    \includegraphics[width=0.4\textwidth]{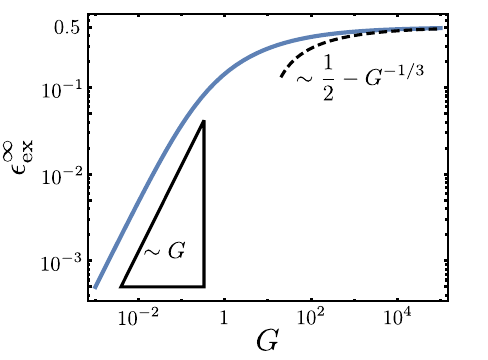}
    \caption{The (dimensionless) excess energy per particle of an infinite stack.  The excess energy goes to zero for stacks with infinitely compliant gaps ($G\to 0$) as particles go to their preferred curvature.  Stacks with infinite gap stiffness ($G\to \infty$) will have conformally stacking particles that have completely flattened meaning the excess energy is just the flattening energy, $\epsilon_{\rm ex}^{\infty} \to 1/2$.  Here the excess energy is normalized, by $Bw\kappa_0^2$ which is twice the flattening energy.}
    \label{fig:einf}
\end{figure}

It is straightforward to solve the above equation by separation of variables to find the curvature $\tilde{\kappa}(h)$ at scaled position $h$ in the stack
\begin{equation}
\label{equation:hinstack}
    h = -\int_{\tilde{\kappa}_-}^{\tilde{\kappa}(h)} \frac{{\rm d}\tilde{\kappa}}{\sqrt{T(\tilde{\kappa},G,\tilde{\kappa}_-)}} \text{ .}
\end{equation}
The total stack size $H$ can similarly be found by integrating over the full range of curvature from $\tilde{\kappa}_-$ to $\tilde{\kappa}_+$ with 
\begin{equation}
\label{equation:stacksizeH}
    H(G, \tilde{\kappa}_-) = -\int_{\tilde{\kappa}_-}^{\tilde{\kappa}_+ = 2 - \tilde{\kappa}_-} \frac{{\rm d}\tilde{\kappa}}{\sqrt{T(\tilde{\kappa},G,\tilde{\kappa}_-)}} \text{ .}
\end{equation}
We note that in this formulation, the stack size depends on the reduced gap stiffness $G$ and is effectively parameterized by the bottom curvature $\tilde{\kappa}_-$, with $H \to 0$ for $\tilde{\kappa}_- \to 1$ for all $G$, and $H \to \infty$ for $\tilde{\kappa}_- \to \tilde{\kappa}^{\rm max}_-(G)$.  This divergence of stack size at a maximal bottom curvature $\tilde{\kappa}^{\rm max}_-$ is shown in more detail in Appendix \ref{appendix:divergingH}.

The excess energy density of an infinite stack can also be calculated.  By taking particle shapes to be uniformly shaped with curvature $\tilde{\kappa}_{\infty}$ at all points in the stack, we can set $\tilde{\kappa}' = 0$ in Eq. (\ref{equation:eexcess}) and find 
\begin{equation}
\label{equation:exinfty}
    \epsilon_{\rm ex}^{\infty}(G) = \frac{1}{2}(\tilde{\kappa}_{\infty}-1)^2
+\frac{G}{2}\tilde{\kappa}_{\infty}^4 \text{ .}
\end{equation}
Finding the curvature which minimizes this energy involves solving for the roots of the equation cubic shown in Eq. (\ref{equation:kinfcubic}).  Doing so, we find only one real root,
\begin{multline}
    \label{equation:kinfty}
    \tilde{\kappa}_{\infty}(G) = \sqrt[3]{\frac{1}{4G}+ \sqrt{\frac{1}{16 G^2}+\frac{1}{216G^3}}} \\+ \sqrt[3]{\frac{1}{4G}- \sqrt{\frac{1}{16 G^2}+\frac{1}{216G^3}}} \text{ .}
\end{multline}
The behavior of $\tilde{\kappa}_{\infty}(G)$ and $\epsilon_{\rm ex}^{\infty}(G)$ are shown in Figs. \ref{fig:kinf} and \ref{fig:einf}, respectively.  We note that for infinitely stiff gaps ($G \to \infty)$, the particles in the infinite stack become completely flattened $\tilde{\kappa}_{\infty} \to 0$ and the stack has energy density $\epsilon_{\rm ex}^{\infty} = 1/2$, which corresponds to the (dimensionless) bending energy for a completely flattened curvamer.  Conversely, for infinitely compliant gaps ($G \to 0)$, the particles maintain their preferred shapes with $\tilde{\kappa}_{\infty}\to 1$. In general, stacks with uniform particles at their preferred shape have energy density $\epsilon_{\rm ex}^{\infty} = G/2$ which is the per particle cost associated with stretching inter-particle gaps to the natural gap size $\delta$.  In the case of completely compliant gaps, we see $\epsilon_{\rm ex}^{\infty} \to 0$.

\section{Diverging stack size}
\label{appendix:divergingH}

\begin{figure}
    \centering
    \includegraphics[width=0.45\textwidth]{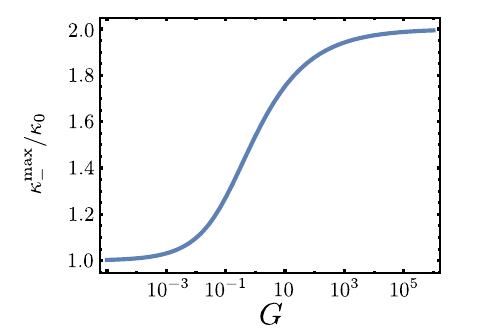}
    \caption{The maximum bottom curvature for a stack with reduced gap stiffness $G$.  Stack size is parameterized by the curvature at the bottom of the stack $\tilde{\kappa}_- \in [1,\tilde{\kappa}_-^{\rm max}]$.  As $\tilde{\kappa}_- \to \tilde{\kappa}_-^{\rm max}$, the stack size continuously diverges $H \to \infty$.}
    \label{fig:kminusmax}
\end{figure}

The squared rate of curvature change function $T({\tilde{\kappa},G,\tilde{\kappa}_-})$ of Eq. (\ref{equation:Tke}) has a minimum at $\tilde{\kappa} = \tilde{\kappa}_{\infty}$.  If this minimum is positive so that $T({\tilde{\kappa},G,\tilde{\kappa}_-})$ has no real roots, then the integrand in Eq. (\ref{equation:stacksizeH}) is well behaved and the stack size can be evaluated.  However, if $T({\tilde{\kappa},G,\tilde{\kappa}_-})$ develops a real root, then the integrand of Eq. (\ref{equation:stacksizeH}) contains a singularity at $\tilde{\kappa} = \tilde{\kappa}_{\infty}$.  To understand how this effects the stack size, we first see if there is choice of the bottom curvature $\tilde{\kappa}_-$ that causes a real root to develop.  Indeed, we find $T(\tilde{\kappa}_{\infty},G,\tilde{\kappa}_-^{\rm max}) = 0$ for   
\begin{align}
    \tilde{\kappa}_-^{\rm max} &= 1 + \sqrt{ (\tilde{\kappa}_{\infty}-1)^2 + G \tilde{\kappa}_{\infty}^4} \nonumber \\ &= 1 + \sqrt{2\epsilon_{\rm ex}^{\infty}(G)} \text{ .}
\end{align}
As seen in Fig. \ref{fig:kminusmax}, in the limit of $G \to \infty$, we see that $\tilde{\kappa}_-^{\rm max} \to 2$, which agrees with the results of Ref. \cite{conformalcurvamers} for conformally contacting, curvature focusing stacks.  For infinitely compliant bonds, $G \to 0$, we see the bottom curvature achieve its preferred shape, $\tilde{\kappa}_-^{\rm max} \to 1$.

We now expand $T({\tilde{\kappa},G,\tilde{\kappa}_-})$ about its minimum to second order and define this approximation as
\begin{equation}
    T^*(\tilde{\kappa},G,\tilde{\kappa}_-) = a(G,\tilde{\kappa}_-) + b(G)(\tilde{\kappa}-\tilde{\kappa}_{\infty})^2\text{ ,}
\end{equation}
with coefficients
\begin{align}
    a(G,\tilde{\kappa}_-) &= T(\tilde{\kappa}_{\infty},G,\tilde{\kappa}_-) \nonumber \\ &= \tilde{\kappa}^4_{\infty} + \frac{1}{G}\Big((\tilde{\kappa}_{\infty}-1)^2 - (\tilde{\kappa}_--1)^2 \Big) \nonumber \\ &=  \frac{1}{G}\Big(( \tilde{\kappa}_-^{\rm max}-1)^2 - (\tilde{\kappa}_--1)^2 \Big)\text{ ,}
\end{align}
and
\begin{align}
    b(G) &= \frac{1}{2}\frac{\partial^2 T(\tilde{\kappa},G,\tilde{\kappa}_-)}{\partial \tilde{\kappa}^2}\bigg|_{\tilde{\kappa}=\tilde{\kappa}_{\infty}} \nonumber \\ &= 6\tilde{\kappa}^2_{\infty} + \frac{2}{G} \text{ .}
\end{align}
The integrand of Eq. (\ref{equation:stacksizeH}) can now be written as
\begin{multline}
    \frac{1}{\sqrt{T(\tilde{\kappa},G,\tilde{\kappa}_-)}} = \frac{1}{\sqrt{T^*(\tilde{\kappa},G,\tilde{\kappa}_-)}} \\ + \left( \frac{1}{\sqrt{T(\tilde{\kappa},G,\tilde{\kappa}_-)}} -  \frac{1}{\sqrt{T^*(\tilde{\kappa},G,\tilde{\kappa}_-)}} \right)\text{ ,} 
\end{multline}
so that the stack size integral can be separated into into two terms
\begin{equation}
    H(G,\tilde{\kappa}_-) = H_1(G,\tilde{\kappa}_-) + H_2(G,\tilde{\kappa}_-) \text{ .}
\end{equation}
Here we define the first term as 
\begin{equation}
\label{equation:H1}
    H_1(G,\tilde{\kappa}_-) = -\int_{\tilde{\kappa}_-}^{\tilde{\kappa}_+} \frac{{\rm d}\tilde{\kappa}}{\sqrt{T^*(\tilde{\kappa},G,\tilde{\kappa}_-)}}\text{ ,}
\end{equation}
and the second term as
\begin{multline}
\label{equation:H2}
    H_2(G,\tilde{\kappa}_-) = \\ -\int_{\tilde{\kappa}_-}^{\tilde{\kappa}_+}\left( \frac{1}{\sqrt{T(\tilde{\kappa},G,\tilde{\kappa}_-)}} -  \frac{1}{\sqrt{T^*(\tilde{\kappa},G,\tilde{\kappa}_-)}} \right){\rm d}\tilde{\kappa} \text{ .}
\end{multline}
Notably, as as $\tilde{\kappa}_- \to \tilde{\kappa}_-^{\rm max}$, $ a(G,\tilde{\kappa}_-) \to 0$ and $T^*(\tilde{\kappa},G,\tilde{\kappa}_-)$ has a single real root at $\tilde{\kappa} = \tilde{\kappa}_{\infty}$, and hence the integrand in $H_1(G,\tilde{\kappa}_-) $ singular, while the integrand in $H_2(G,\tilde{\kappa}_-)$ is finite in this limit.

We are now interested in what happens to these two terms as $\tilde{\kappa}_- \to \tilde{\kappa}^{\rm max}_-$.  For $H_1$, first we evaluate the integral in Eq. (\ref{equation:H1}) and simplify with an addition formula obtaining
\begin{equation}
\label{eq: H1}
    H_1(G,\tilde{\kappa}_-)
    = \frac{1}{\sqrt{b(G)}} {\rm arcsinh}\left[q(G,\tilde{\kappa}_-)\,u(G,\tilde{\kappa}_-)\right]\text{ ,}
\end{equation}
where we've defined
\begin{equation}
\label{eq: q}
    q(G,\tilde{\kappa}_-) = \left(\frac{b(G)}{a(G,
    \tilde{\kappa}_-)}\right) ^{1/2}\text{ ,}
\end{equation}
and 
\begin{multline}
    u(G,\tilde{\kappa}_-) = \sqrt{x_-^2+q^2(G,\tilde{\kappa}_-)x_-^2x_+^2}\\ -\sqrt{x_+^2+q^2(G,\tilde{\kappa}_-)x_-^2x_+^2}\text{ ,}
\end{multline}
with
\begin{align}
    &x_- = \tilde{\kappa}_- - \tilde{\kappa}_{\infty} \text{ ,} \\
    &x_+ = \tilde{\kappa}_+ - \tilde{\kappa}_{\infty}\text{ .}
\end{align}
We note that the bottom curvature is always greater than the top curvature $\tilde{\kappa}_- >\tilde{\kappa}_+ = 2-\tilde{\kappa}_-$ for stacks of size $H>0$, which implies $u(G,\tilde{\kappa}_-) > 0$.  Only in the case of stacks of zero size do we have $\tilde{\kappa}_- = \tilde{\kappa}_+ = 1$, which would imply $u(G,\tilde{\kappa}_-) = 0$.  By taking $\tilde{\kappa}_- \to \tilde{\kappa}_-^{\rm max}(G)$, we see $q(G,\tilde{\kappa}_-) \to \infty$ since $ T(\tilde{\kappa}_{\infty},G,\tilde{\kappa}_-^{\rm max}) = 0$ by definition.  Thus, we see that $H_1(G,\tilde{\kappa}_-)$ continuously diverges as $\tilde{\kappa}_- \to \tilde{\kappa}_-^{\rm max}$.  Finally, we note that by construction, $H_2(G,\tilde{\kappa}_-)$ has no singularity and integrating over the finite range $\tilde{\kappa} \in [2-\tilde{\kappa}_-^{\rm max}, \tilde{\kappa}_-^{\rm max}]$ means that $H_2$ will always be finite in value.  Thus we conclude that the mechanically equilibrated stack size $H(G,\tilde{\kappa}_-) $ of Eq. (\ref{equation:stacksizeH}) continuously diverges as the bottom curvature approaches a maximal value $\tilde{\kappa}_- \to \tilde{\kappa}_-^{\rm max}(G)$, which depends on the dimensionless ratio of inter-particle to intra-particle stiffness.

\section{Excess energy and self-limitation with finite gap stiffness}
\label{appendix:selflimiting-appendix}

The self-limiting stack size $H_*$ is the size which minimizes the total energy density
\begin{equation}
\label{equation:etotal}
    \epsilon = - \epsilon + \frac{S}{H} + \frac{E_{\rm ex}}{H}\text{ ,}
\end{equation}
where we've defined the total (dimensionless) excess energy as $E_{\rm ex}[\tilde{\kappa}(h)] = H\cdot  \epsilon_{\rm ex}[\tilde{\kappa}(h)]$.  Therefore we can obtain an equation of state that describes self-limitation by taking $\frac{\partial \epsilon}{\partial H} =0$ and find 
\begin{equation}
    S (H_*)= H_*^2\frac{d\epsilon_{\rm ex}}{dH} \bigg|_{H=H_*} \text{.}
\end{equation}
However, in Appendix \ref{appendix:curvprofile} we showed that the stack size $H(G,\tilde{\kappa}_-)$ is parameterized by the bottom curvature of the stack $\tilde{\kappa}_-$.  Thus by taking $\frac{\partial \epsilon}{\partial \tilde{\kappa}_-}=0$ we can find a form of the equation of state which is easier to evaluate,
\begin{equation}
    S = H \frac{ \partial E_{\rm ex}}{\partial\tilde{\kappa}_- } \Big(\frac{\partial H}{\partial\tilde{\kappa}_- }\Big)^{-1} - E_{\rm ex} \text{ .}
\end{equation}
For a given stack size $H(G,\tilde{\kappa}_-)$ and reduced gap stiffness $G$, this gives us the reduced cohesion $S$ that makes $H^*=H(G,\tilde{\kappa}_-)$ the minimum of Eq. (\ref{equation:etotal}), a.k.a the self-limiting stack size. 

The excess energy can be simplified to a form that can be evaluated directly. By starting from Eq. (\ref{equation:eexcess}) and utilizing Eq. (\ref{equation:Tke}), it is straightforward to find
\begin{multline}
\label{equation:Eex}
    E_{\rm ex}(G,\tilde{\kappa}_-) = -G\int_{\tilde{\kappa}_-}^{\tilde{\kappa_+}}\sqrt{T(\tilde{\kappa},G,\tilde{\kappa}_-)}{\rm d}\tilde{\kappa} \\ + \frac{H}{2}(\tilde{\kappa}_- - 1)^2 + \frac{G}{3}(\tilde{\kappa}^3_+ -\tilde{\kappa}^3_-)\text{ .}
\end{multline}
To calculate the derivative of the excess energy with respect to $\tilde{\kappa}_-$, we make use of the Leibniz integral rule and find
\begin{equation}
\label{equation:Eexprime}
    E'_{\rm ex} = \frac{H'}{2}(\tilde{\kappa}_- - 1)^2 \text{ .}
\end{equation}
Substituting Eqs. (\ref{equation:Eex}) and (\ref{equation:Eexprime}), we now find
\begin{multline}
    \label{equation:correspondingS}
    S_{\rm SLA}(G,\tilde{\kappa}_-) = G\int_{\tilde{\kappa}_-}^{\tilde{\kappa_+}}\sqrt{T(\tilde{\kappa},G,\tilde{\kappa}_-)}{\rm d}\tilde{\kappa} \\+ \frac{G}{3}(\tilde{\kappa}^3_- - \tilde{\kappa}^3_+) \text{ ,}
\end{multline}
where we denote $S_{\rm SLA}(G,\tilde{\kappa}_-)$ to be the function that returns the value of reduced cohesion which makes $H(G,\tilde{\kappa}_-)$ the self-limiting stack size.

As shown in Appendix \ref{appendix:divergingH}, stack size diverges at a maximum bottom curvature $\tilde{\kappa}_-^{\rm max}(G)$.  We can see what value of $S$ corresponds to when $H_* \to \infty$ by taking $\tilde{\kappa}_-\to \tilde{\kappa}_-^{\infty}(G)$ in Eq. (\ref{equation:correspondingS})   
\begin{equation}
\label{equation:smaxg}
    S_{\rm max}(G) = S_{\rm SLA}(G,\tilde{\kappa}^{\rm max}_-) \text{ .}
\end{equation}
Alternatively, following the results of Section III.B.1 of Ref. \cite{grason-slareview-2021} for limits of self-limitation, we can define the maximal cohesion as 
\begin{equation}
    \label{equation:smaxaccumulant}
    S_{\rm max}(G) = \lim_{ H_{*} \to \infty}{ H_{*} \left[ \epsilon_{\rm ex}^{\infty}(G) - \epsilon_{\rm ex} (G, H_{*} ) \right]} \text{ .}
\end{equation}

\begin{figure}
    \centering
    \includegraphics{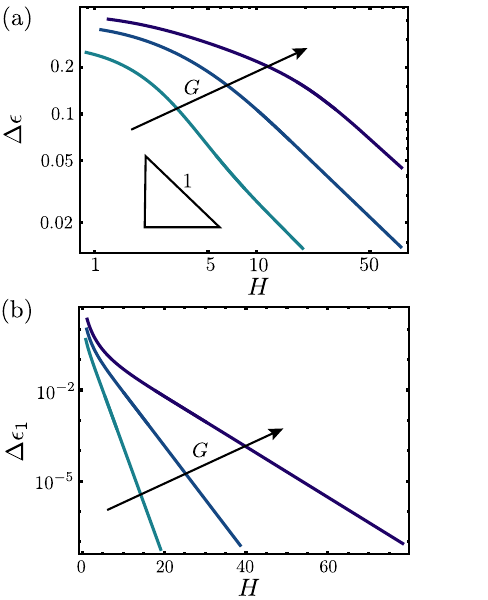}
    \caption{(a)  The residual energy $\Delta \epsilon = \epsilon_{\rm ex}^{\infty} - \epsilon_{\rm ex}(H)$ decreases as $\sim 1/H$ in the limit of large $H$.  (b)  The residual energy $\Delta \epsilon_1(H) = \epsilon_{\rm ex}(H) - \epsilon_{\rm ex}^{\infty} + \frac{S_{\rm max}}{H}$ decreases exponentially in the limit of large $H$.  Curves shown correspond to values of $G = 10$, $100$, $1000$.}
    \label{fig:hdivergence}
\end{figure}

We are now interested in how the self-limiting stack size $H_*$ diverges as $S\to S_{\rm max}$.  For large $H_*$, we know that the excess energy will be near the infinite energy $\epsilon_{\rm ex}^{\infty}$.    Following Ref. \cite{grason-slareview-2021}, we assume that the residual energy 
\begin{equation}
\Delta \epsilon(H_*) = \epsilon_{\rm ex}^{\infty} - \epsilon_{\rm ex}(H_*) \text{ ,}
\end{equation}
vanishes with $H_*$ according to a power law.  Plotting the residual energy in Fig. \ref{fig:hdivergence}a, we find the excess energy density can be approximated as
\begin{equation} 
\label{equation:excesscorrection1}
\epsilon_{\rm ex}(H_*) \simeq \epsilon_{\rm ex}^{\infty}(G) - \frac{C_1(G)}{H_*} \text{ .}
\end{equation}
Solving for $C_1(G)$ in the limit of large $H_*$ yields
\begin{equation}
    C_1(G) = S_{\rm max}(G)\text{ ,}
\end{equation}
according to Eq. (\ref{equation:smaxaccumulant}).  Next, we consider a higher order correction to the excess energy in the limit of large $H_*$.  We define a new residual energy to be the difference between the excess energy and Eq. (\ref{equation:excesscorrection1}), 
\begin{equation}
    \Delta \epsilon_1(H_*) = \epsilon_{\rm ex}(H_*) - \epsilon_{\rm ex}^{\infty} + \frac{S_{\rm max}}{H_*} \text{ ,}
\end{equation}
 which, plotted in Fig. \ref{fig:hdivergence}b, reveals $\Delta \epsilon_1(H_*) \sim e^{-m(G)H_*}$ for some constant $m(G)$ that depends on $G$.  We can now write the total energy density in the limit of large $H_*$ as
\begin{equation}
    \epsilon(H_*) \simeq -\epsilon + \epsilon_{\rm ex}^{\infty} - \frac{S_{\rm max} - S}{H_*} + C_2(G)e^{-m(G)H_*}\text{ .}
\end{equation}
Since $H_*$ is the stack size that minizes of the energy density, we see that for reduced cohesion $S$ near $S_{\max}$ the self-limiting stack size diverges as 
\begin{equation}
    H_* \sim -\ln(S_{max}-S)\text{ .}
\end{equation}

\section{Self limitation with finite interaction range}
\label{appendix:selflimiting-appendix-k}

To recast our theory in terms of finite interaction ranges, we introduced the dimensionless variable $K$ in Eq. (\ref{equation:reducedK}) which quantifies the ratio of surface energy in a stack of height $r_0$ to the cost of cohesive strain induced by the natural gap $\delta$ between particles with their ideal curvature, and depends on a characteristic binding lengthscale $\sigma_{\rm eff} = \sqrt{\gamma/\gamma''}$, or effective interaction range.  This new variable was derived by taking the ratio of $G$ to $S$ to eliminate the particle stiffness $B$ in favor of the purely cohesive variables $\gamma$ and $\gamma''$ which define $\sigma_{\rm eff}$.  Since $K = G/S$, by considering stacks of curvamers with fixed interaction range (and hence fixed $K$) we see that the reduced gap stiffness $G$ must change as the reduced cohesion $S$ is varied. This means that our current methods for calculating stack size using Eq. (\ref{equation:stacksizeH}) and  the reduced cohesion from Eq. (\ref{equation:correspondingS}) must be modified.

\begin{figure}
    \centering
    \includegraphics[width=0.45\textwidth]{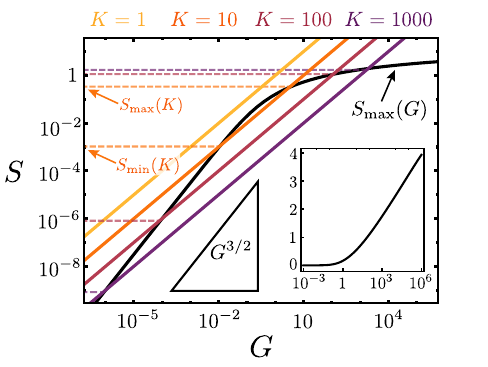}
    \caption{Parameter space of self-limitation.  The self-limiting stack size diverges at maximal cohesion $S_{\rm max}(G)$ (black curve), thus values of $S$ and $G$ below $S_{\rm max}(G)$ represent self-limiting states.  Above $S_{\rm max}(G)$, only unlimited assembly occurs.  Lines of constant $K = G/S$ represent fixed interaction range.  Below a minimal $K^* \simeq 5.0$, self-limitation completely vanishes (yellow line).  For fixed $K>K^*$, there is a finite range of $S$ that permits self-limitation  (between $S_{\rm min}(K)$ and $S_{\rm max}(K)$) which increases with $K$.}
    \label{fig:Smax}
\end{figure}

Additionally, we saw in Appendix \ref{appendix:selflimiting-appendix} there is a maximum value of cohesion $S_{\rm max}(G)$ at which the self-limiting stack size diverges.  In Fig. \ref{fig:Smax}, we plot $S_{\rm max}(G)$ (black curve) against $G$ and note that choices of $S$ and $G$ that lie below $S_{\rm max}(G)$ correspond to self-limiting states, while those on or above the curve are unlimited.  We also see that some lines of constant $K = G/S$ can intersect the $S_{\rm max}$  curve multiple times, and others smaller than a critical value, $K^* \simeq 5$, are completely above the curve.  Since $K\sim \sigma_{\rm eff}^{-2}$, this implies that self-limitation vanishes above some maximum interaction range.  If $K>K^*$,  then these lines intersect twice at the points $S_{\rm min}(K)$ and $S_{\rm max}(K)$ which satisify
\begin{equation}
\label{equation:smaxk}
    S_{\rm max}(KS)=S\text{ ,}
\end{equation}
where we substituted $G = K S$ into Eq. (\ref{equation:smaxg}).  These values represent the bounds of the finite window of cohesive strength that permit self-limitation, and can be seen as the dashed vertical lines in Fig. \ref{fig:computerSLA}.  We numerically solve for the roots of Eq. (\ref{equation:smaxk}) which gives us $S_{\rm min}(K)$ and $S_{\rm max}(K)$.

For a given $K$ and $S\in(S_{\rm min}(K),S_{\rm max}(K))$, we would like to calculate the self-limiting stack size.  To do so, we utilize Eq. (\ref{equation:correspondingS}) and solve for the value of $\tilde{\kappa}_-$ which satisfies
\begin{equation}
    S_{\rm SLA}(KS,\tilde{\kappa}_-) = S \text{ .}
\end{equation}
Again we accomplish this numerically, and plug in this value of the bottom curvature into Eq. (\ref{equation:stacksizeH}) to find the self-limiting stack size $H_* = H(KS,\tilde{\kappa}_-)$.

\section{Measuring gaps in the stack}
To quantify the degree of gap-opened stacking and the stack size at which an assembly transitions from gap-closed to gap-opened stacking, we must calculate the gap size between curvamer surfaces in a stack.  To do this with our continuum model, we take the discrete surface-surface separation of Eq. (\ref{eq: gapstrain}) at the minimal center-center separation distance $\Delta z^* = t + \frac{1}{24}\left(\kappa_{n+1}^- - \kappa_{n}^+ \right)w^2$ in the continuum limit and find the center of the gap ($x=0$) at a height $h$ in the stack to have size
\begin{equation}
    \label{equation:continuumgap}
    \Delta(h) = \frac{t w^2 \kappa_0^2}{24}\left(\tilde{\kappa}_*'(h) + \tilde{\kappa}_*^2(h)\right) \text{ .}
\end{equation}
Consequently, we see that the uniform gap associated with an infinite stack is 
\begin{equation}
    \label{equation:deltainf}
    \delta_{\infty} = \frac{t w^2 \kappa_0^2}{24}\tilde{\kappa}_{\infty}^2(G)\text{ ,} 
\end{equation}
and normalizing the center gap by $\delta_{\infty}$, we obtain
\begin{equation}
    \label{equation:normalizedgap}
    \frac{\Delta(h)}{\delta_{\infty}} = \frac{\tilde{\kappa}_*'(h) + \tilde{\kappa}_*^2(h)}{\tilde{\kappa}_{\infty}(G)} \text{ .}
\end{equation}
We define a stack of size $H$ to be ``gap-opened'' when $\Delta(H/2) =  0.5 \delta_{\infty}$, so therefore must calculate what the curvature is halfway through a stack of a given size.  To do this we specify $G$, and $\tilde{\kappa}_-$ and thus get the stack size $H(G,\tilde{\kappa}_-)$ according to Eq. (\ref{equation:stacksizeH}).  We then numerically solve for the choice of curvature $\tilde{\kappa}(H/2)$ for the upper limit of integration in Eq. (\ref{equation:hinstack}) that makes $h = H/2$.  We can then calculate $\tilde{\kappa}_*'(H/2)$ by substituting $\tilde{\kappa}(H/2)$ into Eq. (\ref{equation:Tke}) and thus find the center gap in the middle of a stack.

Measuring gap distances is more straightforward in the coarse-grained model.  After the conjugate gradient energy minimization protocol has finished for a given stack, we find the positions of the 5 beads in the center of the top surface of the $n^{\rm th}$ curvamer in the stack. Similarly, we find the positions of the 5 beads in the center of the bottom surface of the $n^{\rm th}+1$ curvamer in the stack.  We then calculate the distances between these corresponding beads on the two different curvamers and average together to find the center gap between curvamer surfaces.  This procedure is then performed for the two curvamers in the middle of the stack to obtain $\Delta(H/2)$.  If there are an odd number of curvamers in the stack (and thus an even number of gaps) we perform the above procedure on the two gaps nearest the middle of the stack and average them together.  Finally, to roughly estimate the infinite gap size, we take $\delta_{\infty}$ to be the center gap halfway up a stack of 50 curvamers, which is the largest stack size we consider.

\section{Coarse-grained simulation parameters}
The coarse-grained model we employ to test our continuum theory is adapted from the one introduced by Tanjeem and coworkers in Ref. \cite{conformalcurvamers}, with only a few notable differences.  The first is the geometry of the curvamer which has been scaled down, although is proportionally the same.  For convenience, we set the minimum of all bead-bead interactions (hard core diameter) to $d_{\rm core} = 1$, whereas in Ref. \cite{conformalcurvamers} it was set to $3.55$. We measure all lengths in units of $d_{\rm core}$ and all energies in units of $\varepsilon_0$, where the bead-bead interaction strength is $\varepsilon = \alpha\, \varepsilon_0$.  When measured in units of $d_{\rm core}$ our curvamer design matches that of Ref. \cite{conformalcurvamers}.  We list the geometric parameters used in this article in Table \ref{table:geometry}.  The interaction cutoff distance for the bead-bead potentials is set to $t_0 + 2\, d_{\rm core}$ so that beads in one curvamer only interact with the beads in the neighboring curvamer directly above and below and not next nearest neighbors.  The bead-bead interaction energies are shifted so that the energy is zero at the cutoff distance.  Additionally, interactions between beads in the same curvamer are turned off.  The value of the bead-bead interaction strength $\varepsilon$ is set so that the minimum of the interaction energy between two flat curvamer plates is always $-\gamma w = 1000\, \varepsilon_0$ (See Appendix \ref{appendix:cohesion} for more details). We then make use of the conjugate gradient method in \texttt{LAMMPS} to minimize the energy of a curvamer stack with stopping energy tolerance $10^{-14}$, maximum number of iterations $10^{5}$, maximum number of force/energy evaluations $10^{6}$, and the stopping force tolerance turned off by setting it to zero.  The methods used for measuring the radius of curvature of particles in a stack and for measuring the bending modulus are the same as those used in Ref. \cite{conformalcurvamers}.

\begin{table}
\caption{Structural thickness, width, attractive patch width and preferred radius of curvature of coarse-grained curvamers.}
\begin{tabularx}{\columnwidth}{  >{\centering\arraybackslash}X  >{\centering\arraybackslash}X 
 }
\hline \hline
 Parameter & Value  \\ 
 \hline
 $t_0$ & $1.4$ \\
 $w$ & $13.27$ \\
 $l$ & $4.423$ \\
 $r_0$ & $8.45$ \\
 \hline \hline
\end{tabularx}
\label{table:geometry}
\end{table}

\begin{table}
\caption{Values of interaction range, spring constant, bending modulus, reduced cohesion and gap stiffness used in Fig. \ref{fig:curvatureprofile} and Fig. \ref{fig:gapopening}a.}
\begin{tabularx}{\columnwidth}{  >{\centering\arraybackslash}X >{\centering\arraybackslash}X  >{\centering\arraybackslash}X
>{\centering\arraybackslash}X
>{\centering\arraybackslash}X
 }
\hline \hline 
 $\sigma_{\rm eff}/\delta$ & $k_h$ & $Bw$ & $S$ & $G$  \\ 
 \hline
 $0.13$ & $50\times 10^3$ & 57911 & 0.343 & 8.263 \\
 $0.13$ & $280\times 10^3$ & 324402 & 0.061 & 1.476 \\
 \hline \hline
\end{tabularx}
\label{table:fig4-fig6a}
\end{table}

\begin{table}
\caption{Values of spring constant, bending modulus, reduced cohesion and gap stiffness used in Fig. \ref{fig:computerSLA}a}
\begin{tabularx}{\columnwidth}{  >{\centering\arraybackslash}X  >{\centering\arraybackslash}X
>{\centering\arraybackslash}X
>{\centering\arraybackslash}X
 }
\hline \hline
 $k_h$ & $Bw$ & $S$ & $G$ \\ 
 \hline
 $200\times 10^3$  & 231645 & 0.086 & 7.092 \\
 $160\times 10^3$ & 185316 & 0.108 & 8.906 \\
 $120\times 10^3$ & 138987 & 0.144 & 11.875 \\
 $80\times 10^3$ & 92658 & 0.216 & 13.194 \\
 $60\times 10^3$ & 69493 & 0.288 & 23.749 \\
 $40\times 10^3$ & 46329 & 0.432 & 35.624 \\
 $28\times 10^3$ & 32430 & 0.617 & 50.879 \\
 $24\times 10^3$ & 27797 & 0.720 & 59.373 \\
 $20\times 10^3$ & 23164 & 0.864 & 71.247 \\
 \hline \hline
\end{tabularx}
\label{table:fig9a}
\end{table}

The values of the coarse-grained variables used in Figs. \ref{fig:curvatureprofile} and \ref{fig:gapopening}a are listed in Table \ref{table:fig4-fig6a} along with the continuum model counterparts.  Those used in Fig. \ref{fig:computerSLA}a are listed in Table \ref{table:fig9a}.  Table \ref{table:simparams} lists a mapping between the coarse-grained interaction range $\sigma$ and strength $\varepsilon$ to various continuum model parameters used in the creation of Fig. \ref{fig:phasediagram}b, and applies to all instances of the coarse-grained model mentioned throughout this article. 

\begin{table}[h]
\caption{Mapping from the coarse-grained model to continuum model parameters used in Fig. \ref{fig:phasediagram}b. and throughout this article.}
\begin{tabularx}{\columnwidth}{  >{\centering\arraybackslash}X  >{\centering\arraybackslash}X >{\centering\arraybackslash}X >{\centering\arraybackslash}X >{\centering\arraybackslash}X >{\centering\arraybackslash}X 
 }
\hline \hline
 $\sigma$ &  $\varepsilon$ & $\sigma_{\rm eff}/\delta$ & $t$ & $\gamma''w$ & $K$ \\ 
 \hline
2.0 & 0.874 & 0.145 & 2.343 & 116413 & 19.467 \\
1.95 & 0.881 & 0.141 & 2.344 & 122441 & 20.4861 \\
1.9 & 0.889 & 0.136 & 2.345 & 129178 & 21.624 \\
1.8 & 0.906 & 0.130 & 2.348 & 143940 & 24.122 \\
1.7 & 0.925 & 0.123 & 2.350 & 161772 & 27.139 \\
1.6 & 0.948 & 0.115 & 2.353 & 182951 & 30.726 \\
1.5 & 0.973 & 0.108 & 2.355 & 209261 & 35.182 \\
1.4 & 1.001 & 0.100 & 2.358 & 241867 & 40.709 \\
1.3 & 1.034 & 0.093 & 2.360 & 283200 & 47.718 \\
1.1 & 1.115 & 0.078 & 2.366 & 402180 & 67.921 \\
1.0 & 1.166 & 0.070 & 2.369 & 487669 & 82.462 \\
0.9 & 1.227 & 0.063 & 2.372 & 604352 & 102.320 \\
0.8 & 1.298 & 0.056 & 2.375 & 771478 & 130.778 \\
0.7 & 1.386 & 0.048 & 2.378 & 1027336 & 174.363 \\
0.6 & 1.495 & 0.041 & 2.381 & 1428495 & 242.747 \\
0.5 & 1.636 & 0.034 & 2.383 & 2089178 & 355.443 \\
0.4 & 1.827 & 0.027 & 2.386 & 3189258 & 543.359 \\
 \hline \hline
\end{tabularx}
\label{table:simparams}
\end{table}

\section{Measuring cohesion between two flat plates}
\label{appendix:cohesion}

\begin{figure}
    \centering
    \includegraphics{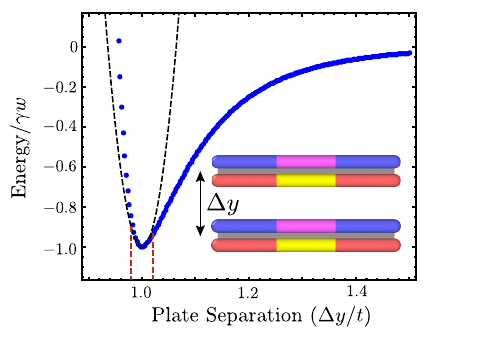}
    \caption{The cohesive energy between two flattened coarse-grained curvamers as calculated in simulations as a function of center-center plate separation.  The dashed black line represents the harmonic approximation $-\gamma w + \frac{1}{2}\gamma''w (\Delta y - t)^2$ at the minimum of the potential well, which was determined by performing a parabolic fit to points in the range $t-0.02\sigma$ to $t+0.02\sigma$ (red dashed lines).  The blue dots shown are for $\sigma_{\rm eff}/\delta = 0.07$.}
    \label{fig:simcohesion}
\end{figure}

To measure the cohesive potential in the coarse-grained model for a particular choice of the bead-bead interaction range $\sigma$, we initialize two flat curvamers with a separation distance $\Delta y$ between the curvamer mid-lines.  We then perform an energy minimization of this configuration with the constraint that all the forces on the beads are set to zero.  This allows \texttt{LAMMPS} to sum over the bead-bead interactions between curvamers and calculate the total energy for this specific configuration in only one step as the minimized state is trivially the initial configuration.  Crucially, we set the preferred curvature and spring constants to zero so that the energy measured is exclusively due to cohesive interactions.  In Fig. \ref{fig:simcohesion}, we show the measured cohesive energy for a range of mid-line separation distances $\Delta y$.  We utilize a golden-section search to locate the minimum of the potential which represents the curvamer thickness $t$ to an accuracy of $10^{-3}$.  The value of the plate-plate interaction $-\gamma w = 1000\,\varepsilon_0$ at its minimum  is kept constant for all choices of the bead-bead interaction range $\sigma$, by choosing the appropriate bead-bead strength $\varepsilon$. 

Mapping from bead-bead potential variables  $\varepsilon$ and $\sigma$ to our dimensionless parameters $G$ and $K$, requires the second derivative of the plate-plate interaction at the minimum, $\gamma''w$.  To do these we fit a parabola to the interaction energy near the minimum of the well with twenty evenly spaced separation distances between $t-0.02\sigma$ and $t+0.02\sigma$ to obtain a value of $\gamma''w$.  As shown in Fig. \ref{fig:simcohesion}, this provides a good harmonic approximation (black dashed line) of the plate-plate interaction near the minimum.

\newpage

\bibliographystyle{apsrev4-2}
\bibliography{main}

\end{document}